\documentclass[aps,prl,reprint,preprintnumbers,superscriptaddress,nofootinbib,longbibliography,floatfix]{revtex4-2}
\usepackage[T1]{fontenc}
\usepackage{lmodern}
\usepackage{amsmath,amssymb,graphicx,booktabs}
\usepackage{microtype}
\usepackage{placeins}
\usepackage{xcolor}
\usepackage[hidelinks]{hyperref}
\hypersetup{pdftitle={Entanglement Requires Fluctuations at Conformal Interfaces},pdfauthor={Yuya Kusuki; Kosei Suzuyama; Yin Tang}}
\newcommand{\ceff}{c_{\mathrm{eff}}}
\newcommand{\ccap}{c_{\mathrm{cap}}}
\newcommand{\Tr}{\operatorname{Tr}}
\newcommand{\dd}{\mathrm{d}}
\newcommand{\sech}{\operatorname{sech}}
\newcommand{\arcosh}{\operatorname{arcosh}}
\newcommand{\arsinh}{\operatorname{arsinh}}

\begin{document}
\preprint{KYUSHU-HET-374}
\preprint{RIKEN-iTHEMS-Report-26}
\title{Entanglement Requires Fluctuations at Conformal Interfaces}
\author{Yuya Kusuki}
\affiliation{\it Institute for Advanced Study, Kyushu University, Fukuoka 812-8581, Japan}
\affiliation{\it Department of Physics, Kyushu University, Fukuoka 819-0395, Japan}
\affiliation{\it RIKEN Center for Interdisciplinary Theoretical and Mathematical Sciences (iTHEMS), Wako, Saitama 351-0198, Japan}
\author{Kosei Suzuyama}
\affiliation{\it Department of Physics, Kyushu University, Fukuoka 819-0395, Japan}
\author{Yin Tang}
\affiliation{\it Institute for Advanced Study, Kyushu University, Fukuoka 812-8581, Japan}
\begin{abstract}
We conjecture that the universal logarithmic coefficient for capacity of entanglement $\ccap$ is bounded below by the corresponding entropy coefficient $\ceff$ at conformal interfaces. We prove the bound in free boson and fermion CFTs and in holographic models with interface branes, and establish it using defect perturbation theory. Numerical calculations at interacting interfaces between the tricritical Ising and Ising CFTs further support the bound. Despite this constraint, capacity provides independent interface data. We construct interfaces with identical entropy and energy transmission coefficients but different capacity coefficients, and show that capacity can even increase as entropy decreases.
\end{abstract}
\maketitle

\paragraph*{Introduction.---}Energy transmission and entanglement entropy are central observables of conformal interfaces, characterizing energy transport and quantum correlations between critical systems~\cite{Oshikawa1996,Oshikawa1997,Bachas2002,Quella2007,Sakai2008,Eisler2010,Peschel2012,Calabrese2012,Brehm2015,Gutperle2016Topological,Gutperle2016Holographic,Wen2018,Rogerson2022,Tang2024,Meineri2020,Bachas2020,Bachas2023Thick,Liu2025,Barad2025,KarchWang2026,Barad2026}. Entanglement entropy gives the mean of the modular Hamiltonian, while its variance, quantified by the \emph{capacity of entanglement}, probes an additional feature of the entanglement spectrum. We show that capacity provides independent universal interface data: interfaces with identical energy transmission and entropy can have different capacities. We further conjecture that the capacity coefficient is universally bounded below by the entropy coefficient.

For the reduced density matrix $\rho_A$ and modular Hamiltonian $K_A=-\log\rho_A$, the entropy and capacity are~\cite{Yao2010,Nakaguchi2016,deBoer2019,Kawabata2021,Nandy2021,Arias2023}:
\begin{equation}
S_A=\langle K_A\rangle,\qquad C_A=\langle K_A^2\rangle-\langle K_A\rangle^2.
\label{eq:definitions}
\end{equation}
The entanglement energies $-\log p_i$ occur with probabilities $p_i$, the eigenvalues of $\rho_A$: entropy fixes their mean, while capacity measures their spread. For a maximally entangled state of Schmidt rank $d$, all nonzero $p_i=1/d$, giving $S_A=\log d$ but $C_A=0$.

Capacity also controls how efficiently entanglement can be converted into Bell pairs. For $N$ copies of a bipartite pure state, let $E_{\mathrm{dist}}$ denote the entanglement extractable as Bell pairs and $E_{\mathrm{cost}}$ the entanglement needed to prepare the copies by local operations and classical communication. At fixed error $0<\delta<1/2$, these obey~\cite{Datta2015}
\begin{equation}
\begin{aligned}
E_{\mathrm{dist}}&=NS_A+\sqrt{NC_A}\,\Phi^{-1}(\delta)+O(\log N),\\
E_{\mathrm{cost}}&=NS_A-\sqrt{NC_A}\,\Phi^{-1}(\delta)+O(\log N),
\end{aligned}
\label{eq:finitecopies}
\end{equation}
where $\Phi$ is the standard normal cumulative distribution function. Since $\Phi^{-1}(\delta)<0$, larger capacity lowers the extraction yield and raises the preparation cost relative to $NS_A$. Entropy thus determines the common asymptotic rate, while capacity controls the leading penalty at finite~$N$. The role of capacity at finite resources has also been studied in holography~\cite{Bao2026}.

Consider two unitary two-dimensional CFTs with central charges $c_L,c_R$, joined by a conformal interface at $x=0$. In the ground state, the entropy and capacity of the right half $A$ of a system of size $2l$ scale as
\begin{equation}
\begin{aligned}
S_A&=\frac{\ceff}{6}\log\frac{l}{\epsilon}+\cdots,\\
C_A&=\frac{\ccap}{6}\log\frac{l}{\epsilon}+\cdots,
\end{aligned}
\label{eq:scaling}
\end{equation}
where $\epsilon$ is the short-distance cutoff. We propose the following inequality:
\begin{equation}
\ceff\leq\ccap.
\label{eq:bound}
\end{equation}
The conjecture therefore bounds the leading finite-$N$ corrections in Eq.~\eqref{eq:finitecopies}.

A homogeneous CFT vacuum saturates the proposed bound, with $\ceff=\ccap=c$~\cite{deBoer2019,Calabrese2004}. The known bound $\ceff\leq\min(c_L,c_R)$ limits how much entanglement an interface supports~\cite{Karch2023}. The conjectured inequality $c_{LR}\leq\ceff$ relates this entanglement to energy transmission, encoded in the stress-tensor coefficient $c_{LR}$~\cite{Karch2024}. Equation~\eqref{eq:bound} would constrain the fluctuations that must accompany it.

We prove the bound at arbitrary transmission in free boson and fermion CFTs. Using defect perturbation theory, we show that weak reflection lowers entropy but raises capacity in interacting CFTs, establishing the bound to leading order. In holographic models with interface branes, we establish the bound and construct interfaces with identical entropy coefficients and energy transmission but different capacities. We then verify the free-field results on the lattice and further test the bound at interacting interfaces. Finally, we show that partially transmitting interfaces can also saturate the bound.

Let $Z_n$ be the partition function on the $n$-sheeted replica geometry. The replica trick gives
\begin{equation}
\begin{gathered}
F(n)=\log\Tr\rho_A^n=\log Z_n-n\log Z_1,\\
S_A=-F'(1),\qquad C_A=F''(1).
\end{gathered}
\label{eq:replica}
\end{equation}
The state $\rho_A(n)=\rho_A^n/\Tr\rho_A^n$ is a thermal ensemble for $K_A$ at inverse temperature $n$. Its entropy obeys
\begin{equation}
-\left.\frac{\dd}{\dd n}S\bigl(\rho_A(n)\bigr)\right|_{n=1}=C_A.
\label{eq:capacityresponse}
\end{equation}
Capacity is therefore the heat capacity associated with $K_A$ at unit temperature~\cite{deBoer2019}.

\begin{figure}[tbp]
\centering
\includegraphics[width=0.8\columnwidth]{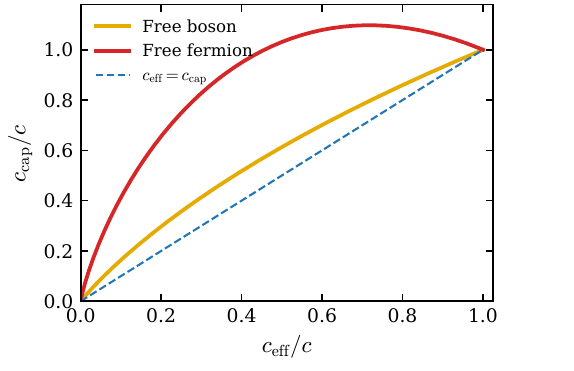}
\caption{The exact free-boson and free-fermion interface coefficients satisfy $\ceff\leq\ccap$ and follow distinct curves. The fermionic capacity can exceed its transparent value $\ccap/c=1$, whereas $\ceff/c\leq1$.}
\label{fig:bound}
\end{figure}

\paragraph*{Free CFTs.---}We consider conformal interfaces for a free boson~\cite{Bachas2002} or a Majorana fermion~\cite{Oshikawa1996,Oshikawa1997}. These interfaces linearly relate incoming and outgoing modes, with transmission coefficient $\mathcal T=s^2$, $0\leq s\leq1$. Figure~\ref{fig:bound} shows $\ccap$ and $\ceff$ obtained from the known expressions for $F(n)$~\cite{Sakai2008,Peschel2012,Brehm2015}. Both curves satisfy $\ceff\leq\ccap$. An analytic proof for all $0\leq s\leq1$ is given in the Supplemental Material~\cite{SM}.

For fermions, capacity can even exceed its transparent value. Near $s=1$,
\begin{equation}
\begin{aligned}
\frac{\ceff}{c}&=1-\frac32(1-s)+O\bigl((1-s)^2\bigr),\\
\frac{\ccap}{c}&=1+(1-s)+O\bigl((1-s)^2\bigr).
\end{aligned}
\label{eq:enhancement}
\end{equation}
Weak reflection therefore lowers entropy but raises capacity. Reflection raises the single-particle entanglement energies and suppresses their occupation. The variance also weights the squared energies, allowing it to increase even as entropy decreases. For bosons both coefficients instead decrease, $\ceff/c=1-\frac34(1-s)+\cdots$ and $\ccap/c=1-\frac12(1-s)+\cdots$, because a bosonic mode at zero entanglement energy already carries the maximal variance. Nevertheless, the bound still holds, with $\ccap-\ceff=\frac14c(1-s)+\cdots$. The two families obey exact sum rules, $\ceff^{B}+\ceff^{\rm Maj}=\frac32s$ and $\ccap^{B}+\ccap^{\rm Maj}=\frac32s(2-s)$ ~\cite{Peschel2012,SM}.

\paragraph*{Defect perturbation theory.---}The increase of capacity extends beyond free fields. We perturb the identity interface by a normalized spinless primary $\phi$ of dimension $1-\varepsilon$, with $|\varepsilon|\ll1$. Let $g$ denote its dimensionless coupling at a nearby conformal fixed point; $\varepsilon=0$ includes exactly marginal deformations.

The one-point function vanishes, so the leading correction is determined by the two-point function on the replica geometry. We analytically continue the replica sum of Ref.~\cite{Brehm2021} to obtain $\delta F(n)$, the change in $F(n)$ relative to the identity interface:
\begin{equation}
\delta F(n)=2ng^2\log\frac{l}{\epsilon}\int_0^\infty\dd x\,[\coth x\tanh(nx)-1]+\cdots.
\label{eq:pertreplica}
\end{equation}
Its first two derivatives yield
\begin{equation}
\begin{aligned}
\ceff&=c-\frac{3\pi^2}{2}g^2+O(g^3,\varepsilon g^2),\\
\ccap&=c+\pi^2g^2+O(g^3,\varepsilon g^2).
\end{aligned}
\label{eq:pertresult}
\end{equation}
The entropy decreases while the capacity increases.

In terms of the reflection coefficient $\mathcal R=2\pi^2g^2/c+O(g^3,\varepsilon g^2)$~\cite{Brehm2021}, this gives
\begin{equation}
\ccap-\ceff=\frac54c\mathcal R+O(g^3,\varepsilon g^2).
\label{eq:pertgap}
\end{equation}
For several such operators, $g^2$ is replaced by the positive quadratic form of their two-point functions. Unitarity therefore implies $\ceff\leq\ccap$ to leading order.

\paragraph*{Holographic interfaces.---}In holography, entropy is determined by an extremal length~\cite{Ryu2006}, while capacity measures its response to the replica cosmic brane~\cite{Nakaguchi2016,deBoer2019,Dong2016}. Geometries with the same entropy coefficient can therefore have different capacity coefficients.

\begin{figure}[tbp]
\centering
\includegraphics[width=0.7\columnwidth]{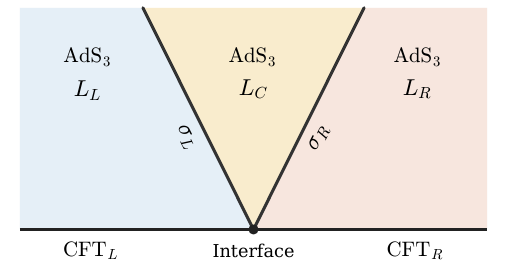}
\caption{Schematic constant-time slice of the double-brane geometry. The branes of tensions $\sigma_L,\sigma_R$ separate AdS$_3$ regions of radii $L_L,L_C,L_R$ and end on the same interface between the two boundary CFTs.}
\label{fig:holographic}
\end{figure}

We implement this calculation at classical order in three-dimensional Einstein gravity with pure-tension interface branes. Two such branes join three AdS$_3$ regions with radii $L_L,L_C,L_R$ and a common Newton constant $G_N$~\cite{Baig2022}. Both branes end on the same boundary interface, while the middle region has finite bulk width (Fig.~\ref{fig:holographic}). The central charges of the two CFTs are $c_{L,R}=3L_{L,R}/(2G_N)$. For $0<L_C<\min(L_L,L_R)$, the Israel junction conditions~\cite{Israel1966} for matching the geometries across each brane restrict the tensions to
\begin{equation}
\frac1{L_C}-\frac1{L_i}<8\pi G_N\sigma_i<\frac1{L_C}+\frac1{L_i},\quad i=L,R.
\label{eq:doubletension}
\end{equation}
The minimum warp factor determines the entropy~\cite{Karch2023,Karch2024,Karch2021,KarchWang2023}, while our calculation of the linearized replica geometry gives the capacity~\cite{SM}:
\begin{equation}
\begin{aligned}
\ceff&=\frac{3L_C}{2G_N},\\
\ccap&=\frac{2L_C\,\ceff}{\displaystyle\frac{L_L}{1+8\pi G_NL_L\sigma_L}+\displaystyle\frac{L_R}{1+8\pi G_NL_R\sigma_R}}.
\end{aligned}
\label{eq:doublecoefficients}
\end{equation}
The lower bound in Eq.~\eqref{eq:doubletension} makes each term in the denominator smaller than $L_C$, proving $\ceff<\ccap$ throughout this family. Equality is approached when both tensions approach their lower endpoints. Thus the Israel junction conditions enforce the proposed bound.

The capacity coefficient remains independent even when the entropy and energy transmission are fixed. At fixed $L_C$, changing the tensions leaves $\ceff$ unchanged. The coefficient $c_{LR}$ is~\cite{Baig2022}
\begin{equation}
c_{LR}=\frac{3}{G_N\bigl[L_L^{-1}+L_R^{-1}+8\pi G_N(\sigma_L+\sigma_R)\bigr]}.
\label{eq:doubletransmission}
\end{equation}
At fixed radii it depends only on the total tension. Capacity also depends on how that tension is divided between the two branes.
Table~\ref{tab:independence} gives two interfaces with the same $c_L,c_R$, $\ceff$, and $c_{LR}$ but different $\ccap$. 

\begin{table}[!t]
\caption{Capacity distinguishes interfaces with identical entropy coefficients and energy transmission. Both double-brane models have $L_L=L_R=L$, $L_C=L/2$, and $c=3L/(2G_N)$. Redistributing a fixed total tension preserves $\ceff$ and $c_{LR}$ but changes $\ccap$.}
\label{tab:independence}
\centering
\begin{tabular}{@{}cccc@{}}
\toprule
$8\pi G_NL(\sigma_L,\sigma_R)$ & $\ceff/c$ & $c_{LR}/c$ & $\ccap/c$\\
\midrule
$(2,2)$ & $1/2$ & $1/3$ & $3/4$\\
$(3/2,5/2)$ & $1/2$ & $1/3$ & $35/48$\\
\bottomrule
\end{tabular}
\end{table}

\begin{figure}[tbp]
\centering
\includegraphics[width=0.85\columnwidth]{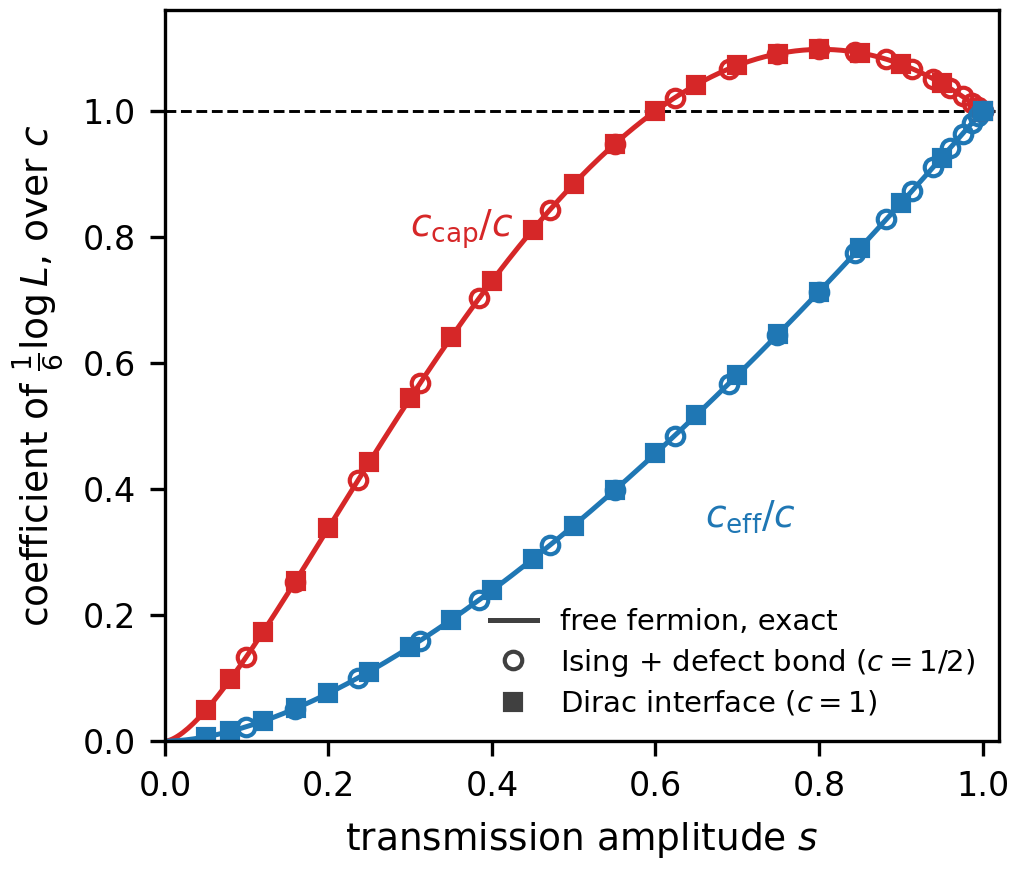}
\caption{Lattice coefficients (symbols), divided by the bulk central charge, for the critical Ising chain with a defect bond of strength $\lambda$ (open circles, $s=2\lambda/(1+\lambda^2)$) and for the Dirac fermion chain with an exactly conformal interface (filled squares, $s=t$). Solid lines are the exact free-fermion coefficients of Fig.~\ref{fig:bound}. Both lattice models follow the fermionic curves.}
\label{fig:lattice}
\end{figure}

\paragraph*{Lattice verification.---}We first test the free-fermion prediction on two lattice realizations of the Majorana interface. The first is the critical transverse-field Ising chain, $H=-\sum_j(\sigma^z_j\sigma^z_{j+1}+\sigma^x_j)$, with the central bond multiplied by $\lambda$; this is the lattice form of the Oshikawa--Affleck defect line~\cite{Oshikawa1996,Oshikawa1997}, with transmission amplitude $s=2\lambda/(1+\lambda^2)$~\cite{Eisler2010}. The second is a free Dirac fermion chain, $H=-\sum_jt_j(c^\dagger_jc_{j+1}+\mathrm{h.c.})+\sum_j\mu_jc^\dagger_jc_j$, with homogeneous hopping $t=1$ and potential $\mu=0$, except for the central hopping $t=s$ and on-site potentials $\mu=\pm\sqrt{1-s^2}$ on the two adjacent sites. The potentials make the single-particle scattering matrix independent of energy, so the interface is exactly conformal on the lattice with transmission amplitude $s$~\cite{Peschel2012,Eisler2012Evolution,Capizzi2023}; it decomposes into two identical Majorana sectors and therefore has $c=1$. Both models are Gaussian, so the entanglement spectrum of the half chain $A$ follows from the eigenvalues $\nu_k$ of the correlation matrix restricted to $A$~\cite{Peschel2003}. Each $\nu_k$ is the occupation of an independent two-level mode of $K_A$, whence
\begin{equation}
\begin{aligned}
S_A&=-\sum_k\bigl[\nu_k\log\nu_k+(1-\nu_k)\log(1-\nu_k)\bigr],\\
C_A&=\sum_k\nu_k(1-\nu_k)\log^2\frac{1-\nu_k}{\nu_k}.
\end{aligned}
\label{eq:gaussian}
\end{equation}
We use open chains with the interface at the center, $L\leq2048$ for Ising and $L\leq4096$ for Dirac, and extract $\ceff$ and $\ccap$ from the $\log L$ dependence. The leading lattice correction to $S_A$ is of order $L^{-1}$. Because $C_A=-2\partial_nS^{(n)}_A|_{n=1}$ and the corresponding R\'enyi correction scales as $L^{-1/n}$~\cite{Cardy2010}, the replica derivative acts on the exponent and the capacity acquires an additional $L^{-1}\log L$ term (see detailed discussion in~\cite{SM}). Figure~\ref{fig:lattice} compares $\ceff/c$ and $\ccap/c$ for both models with the exact curves of Fig.~\ref{fig:bound}. They agree with the theoretical prediction for $\ceff$ and $\ccap$ over the full range of $s$. Normalized by $c$, both lattice models collapse onto the fermionic curve.

\begin{figure}[tbp]
\centering
\includegraphics[width=\columnwidth]{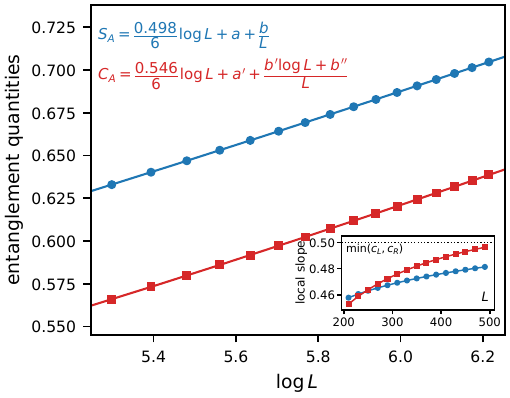}
\caption{Entropy (circles) and capacity (squares) across the $W_3$ interface between the TCI and Ising CFTs, for $200\leq L\leq500$. Solid lines show the reference fit with the forms given, where $b/L$ and $(b'\log L+b'')/L$ are the leading finite-size corrections. Inset: local slopes $6\,\dd S_A/\dd\log L$ (circles) and $6\,\dd C_A/\dd\log L$ (squares). The capacity slope overtakes the entropy slope at $L\simeq240$, and both are still increasing at the largest sizes; the dotted line marks $\min(c_L,c_R)=1/2$.}
\label{fig:W3}
\end{figure}

\paragraph*{Interacting interfaces.---}We next test the bound at interacting interfaces between the tricritical Ising (TCI) and Ising CFTs, with $c_L=7/10$ and $c_R=1/2$. We first consider the recently identified $W_3$-preserving interface of Ref.~\cite{Antunes2026}, realized on the O'Brien--Fendley chain~\cite{OBrien2018}. Its left half is tuned to the TCI point, with couplings rescaled to match the Ising velocity~\cite{Zou2018}; its right half is the critical transverse-field Ising chain; and the two are joined through a single bond of strength $h_b=1.379$~\cite{SM}. After folding, the interface becomes a conformal boundary state of the $c=6/5$ theory that preserves not the product of the two Virasoro algebras but the $W_3$ algebra of the coset $[su(3)_2\oplus su(3)_1]/su(3)_3$, whose additional weights $1/9$, $7/9$ and $13/9$ appear in the interface spectrum~\cite{Antunes2026}. It commutes with the $\mathbb Z_2$ symmetry and with the Kramers--Wannier duality, and supports a single relevant $\mathbb Z_2$-even interface operator, of dimension $\simeq0.5$, which the choice of $h_b$ tunes away. We compute ground states of open chains with $200\leq L\leq500$ using the density-matrix renormalization group~\cite{White1992,Hauschild2018}, and obtain $S_A$ and $C_A$ from the Schmidt spectrum at the interface bond. Fitting $S_A$ and $C_A$ with the finite-size forms of the free-fermion analysis, with leading corrections $L^{-\kappa}$ and $L^{-\kappa}\log L$ (Fig.~\ref{fig:W3}), gives
\begin{equation}
\ceff=0.498(4),\qquad\ccap=0.546(5),
\label{eq:W3}
\end{equation}
where the quoted uncertainties account for fitting-window and correction-form dependence at fixed $\kappa=1$~\cite{SM}. Varying the correction exponent over \(0.8\leq\kappa\leq1.2\) changes \(\ceff\) by at most \(0.007\) and \(\ccap\) by at most \(0.020\), while \(\ccap-\ceff\) remains positive.
The entropy coefficient nearly saturates its upper bound $\min(c_L,c_R)=1/2$. This value is expected from an analytical perspective: the energy transmission coefficient for this interface has been determined to be $c_{LR}\simeq0.498$ ~\cite{Furuta2026}, so the conjectured inequality $c_{LR}\leq\ceff$~\cite{Karch2024} confines $\ceff$ to a window of width $0.002$ below $1/2$, consistent with our value. The capacity coefficient exceeds the entropy by $0.042$--$0.050$ for all these fits, and by at least $0.035$ when the correction exponent is varied as well, in agreement with Eq.~\eqref{eq:bound}. This does not rest on the extrapolation alone: the local slope of $C_A$ overtakes that of $S_A$ at $L\simeq240$, and the difference grows with $L$ (Fig.~\ref{fig:W3}, inset). The fit gives \(\ccap=0.546(5)\), above \(\min(c_L,c_R)=1/2\), although this comparison relies on the finite-size extrapolation. More importantly, \(\ccap-\ceff>0\) is already visible at the level of the local slopes and remains positive under the fitting variations~\cite{SM}.

We have also examined the renormalization-group domain wall between the same two CFTs~\cite{Tang2024,Cogburn2024,Gaiotto2012}, realized by an O'Brien--Fendley chain whose three-site coupling is $g_c=g_{\text{TCI}}\simeq0.428$ on one half and $g_1<g_c$ on the other~\cite{SM}. In the range $0.32\lesssim g_1\lesssim0.38$, where the data are consistent with the RG-domain-wall regime, the fitted coefficients satisfy $\ceff\leq\ccap$. We find $\ccap\simeq0.50$, while $\ceff=0.42$--$0.49$ approaches from below the value $\min(c_L,c_R)=1/2$ expected for this interface~\cite{Tang2024}. These results are thus also consistent with Eq.~\eqref{eq:bound} ~\cite{SM}.

\paragraph*{Saturation of the bound.---}Partially transmitting interfaces can saturate the bound. A topological interface has the same logarithmic term in $F(n)$ as a homogeneous CFT~\cite{Gutperle2016Topological,Brehm2016}, so $\ceff=\ccap=c$, including for noninvertible interfaces in rational CFTs. A factorizing interface has $\ceff=\ccap=0$.

Now consider two tensor-product CFTs sharing a factor of central charge $c_t$. Let the interface be topological on the shared factor and reflecting on the remaining factors. The reduced density matrix factorizes, so entropy and capacity add:
\begin{equation}
\ceff=\ccap=c_t.
\label{eq:saturation}
\end{equation}
For $0<c_t<\min(c_L,c_R)$, this saturates the bound despite partial transmission. The shared factor may itself be an interacting CFT.

\paragraph*{Discussion.---}The conjectured bound links two independent properties of entanglement. Capacity can rise when entropy falls, and its universal coefficient can vary even when the entropy and energy transmission are fixed. Our calculations for free CFTs and holographic interfaces, together with defect perturbation theory and lattice results for free and interacting interfaces, all support $\ceff\leq\ccap$. The conjecture would therefore also set a lower bound on the magnitudes of the leading finite-$N$ corrections to entanglement conversion.

The bound imposes a condition on replica data beyond ordinary R\'enyi monotonicity. Writing $S_A^{(n)}=F(n)/(1-n)$ gives
\begin{equation}
\left.\frac{\partial}{\partial n}\frac{S_A^{(n)}}{1+1/n}\right|_{n=1}=\frac{S_A-C_A}{4}.
\label{eq:renyislope}
\end{equation}
The factor $1+1/n$ removes the leading R\'enyi dependence of a homogeneous CFT. Our conjecture requires the remaining logarithmic coefficient to have a nonpositive slope at $n=1$, whereas ordinary monotonicity gives only $C_A\geq0$.

The central theoretical goal is to prove the bound for general unitary conformal interfaces and identify the principle that enforces a minimum capacity at fixed entropy. The conditions for equality also remain to be understood.
Understanding the mechanism behind this bound may provide clues to more general principles governing entanglement fluctuations. Such insights could ultimately inform the development of entanglement conversion protocols.

In holography, the next step is to extend the brane results to interfaces supported by bulk matter. This requires calculating how both the geometry and the matter fields respond to the replica cosmic brane. Such calculations would test whether the bulk equations enforce $\ceff\leq\ccap$ beyond the pure-tension brane models.

The $W_3$ interface provides a test of the bound beyond free fields and perturbation theory. Together with the free fermion, it also shows that, unlike $\ceff$, the capacity is not limited by $\min(c_L,c_R)$; whether $\ccap$ obeys any upper bound remains open. Our results for the RG domain wall are consistent with the bound as well. Capacity could also distinguish candidate conformal interfaces with the same entropy coefficient, extending the comparison in Table~\ref{tab:independence} to interacting systems.

More broadly, our results show that the capacity carries information about conformal interfaces that is contained in neither $c_{LR}$ nor $\ceff$. An interesting direction is therefore to ask what physical information the capacity of entanglement captures in other quantum many-body systems. The capacity is only the second cumulant of $K_A$: the full replica dependence of the logarithmic coefficient in Eq.~\eqref{eq:renyislope}, equivalently the higher cumulants of the entanglement spectrum, may provide further independent interface data and could obey constraints of its own. Additionally, after a quantum quench, capacity and entropy grow at different rates even in homogeneous free chains~\cite{Arias2023}, and the way an interface filters this growth may reveal information beyond entanglement transport. Finally, entanglement Hamiltonians and their spectra can now be reconstructed in trapped-ion quantum simulators~\cite{Kokail2021,Joshi2023}, and a Rydberg-atom realization of the $W_3$ interface has been proposed~\cite{Antunes2026}. The capacity of entanglement across interfaces may thus become accessible experimentally.

\begin{acknowledgments}
\paragraph*{Acknowledgments.---}YK is supported by the INAMORI Frontier Program at Kyushu University, JSPS KAKENHI Grant Number 23K20046, MEXT Supporting Pioneering Research through AI for 1,000 Discovery challenges Program (SPReAD) Japan Grant Number JPMXP1726262823, and JST PRESTO Grant Number JPMJPR26A6.
We used ChatGPT (OpenAI, GPT-6 astra) and Claude (Anthropic, Opus 5) to assist with drafting the manuscript and writing the numerical codes under our direction. All AI-assisted content was reviewed and verified by the authors.
\end{acknowledgments}

\onecolumngrid
\setcounter{equation}{0}
\setcounter{section}{0}
\setcounter{subsection}{0}
\setcounter{table}{0}
\setcounter{figure}{0}
\setcounter{secnumdepth}{2}
\renewcommand{\theequation}{S\arabic{equation}}
\renewcommand{\thesection}{S\arabic{section}}
\renewcommand{\thetable}{S\arabic{table}}
\renewcommand{\thefigure}{S\arabic{figure}}
\renewcommand{\theHequation}{supplement.\arabic{equation}}
\renewcommand{\theHsection}{supplement.\arabic{section}}
\renewcommand{\theHtable}{supplement.\arabic{table}}
\renewcommand{\theHfigure}{supplement.\arabic{figure}}
\section*{Supplemental Material}
\label{sec:supplement}

\section{Conventions}

We consider the ground state of two unitary two-dimensional CFTs with central charges $c_L,c_R$, joined by a conformal interface at $x=0$. The subsystem $A$ is the right half of a system of size $2l$, and $\epsilon$ is the short-distance cutoff. Let $\rho_A$ be the normalized reduced density matrix of $A$ and $K_A=-\log\rho_A$ its modular Hamiltonian.

Let $Z_n$ denote the partition function on the replica geometry obtained by cyclically gluing $n$ copies along $A$. We define
\begin{equation}
F(n)=\log\Tr\rho_A^n=\log Z_n-n\log Z_1.
\label{sm:eq:replicadefinition}
\end{equation}
The entropy and capacity are
\begin{equation}
S_A=-F'(1)=\Tr(\rho_AK_A),\qquad C_A=F''(1)=\Tr(\rho_AK_A^2)-S_A^2.
\label{sm:eq:entropycapacity}
\end{equation}
Their universal coefficients are defined by
\begin{equation}
S_A=\frac{\ceff}{6}\log\frac{l}{\epsilon}+\cdots,\qquad C_A=\frac{\ccap}{6}\log\frac{l}{\epsilon}+\cdots.
\label{sm:eq:coefficientnormalization}
\end{equation}
The omitted terms are subleading as $l/\epsilon\to\infty$.

All logarithms are natural, so entanglement is measured in nats; division by $\log 2$ gives the number of Bell pairs. For entanglement distillation and formation, we use squared-fidelity error $0<\delta<1/2$: the output state $\rho_{\mathrm{out}}$ must satisfy $\langle\mathrm{target}|\rho_{\mathrm{out}}|\mathrm{target}\rangle\geq1-\delta$, where $|\mathrm{target}\rangle$ is the desired pure output state. The many-copy limit is taken at fixed finite-dimensional regulator before the large-system limit.

\section{Free-field interfaces}

We treat the Majorana fermion and the free boson, which are different conformal interfaces, and note along the way that the Dirac fermion follows from the Majorana case by a factor of two.

\subsection{Majorana fermion, free boson, and Dirac fermion}

For a real boson ($c=1$) or a Majorana fermion ($c=1/2$), $F(n)$ is known explicitly~\cite{Sakai2008,Peschel2012,Brehm2015}:
\begin{align}
F_B(n)&=\frac{\log\frac{l}{\epsilon}}{\pi^2}\int_0^\infty\dd x\left[n\log(1-e^{-2\omega_B})-\log(1-e^{-2n\omega_B})\right]+\cdots,\notag\\
F_F(n)&=\frac{\log\frac{l}{\epsilon}}{\pi^2}\int_0^\infty\dd x\left[\log(1+e^{-2n\omega_F})-n\log(1+e^{-2\omega_F})\right]+\cdots,
\label{sm:eq:realreplica}
\end{align}
where $\omega_B=\arsinh(\sinh x/s)$, $\omega_F=\arcosh(\cosh x/s)$, and $\mathcal T=s^2$ is the transmission coefficient. These expressions hold for real $n>0$ and $0<s\leq1$. The single-particle entanglement energies are $2\omega_{B,F}$; the fermionic spectrum is gapped, $2\omega_F\geq2\arcosh(1/s)>0$, while the bosonic one is not, since $\omega_B(0)=0$~\cite{Peschel2012}. Taking two derivatives at $n=1$ immediately gives
\begin{equation}
\ccap^B=\frac{6}{\pi^2}\int_0^\infty\frac{\omega_B^2}{\sinh^2\omega_B}\,\dd x,\qquad \ccap^F=\frac{6}{\pi^2}\int_0^\infty\frac{\omega_F^2}{\cosh^2\omega_F}\,\dd x.
\label{sm:eq:freecapacities}
\end{equation}
The entropy coefficients follow from the first derivative. Integrating those expressions by parts, with vanishing boundary terms, gives
\begin{equation}
\begin{aligned}
\ccap^B-\ceff^B&=\frac{6}{\pi^2}\int_0^\infty\frac{\omega_B(\omega_B-x\partial_x\omega_B)}{\sinh^2\omega_B}\,\dd x\geq0,\\
\ccap^F-\ceff^F&=\frac{6}{\pi^2}\int_0^\infty\frac{\omega_F(\omega_F-x\partial_x\omega_F)}{\cosh^2\omega_F}\,\dd x\geq0.
\end{aligned}
\label{sm:eq:positivegap}
\end{equation}
The bosonic integrand is nonnegative because $\omega_B(x)$ is concave with $\omega_B(0)=0$. For fermions, $\omega_F\geq x$ and $0\leq\partial_x\omega_F=\sinh x/\sqrt{\sinh^2x+1-s^2}\leq1$ give the same conclusion. For both bosons and fermions, the inequality is strict for $0<s<1$. At $s=1$, $\omega_B=\omega_F=x$ and $\ccap=\ceff=c$; at $s=0$, both coefficients vanish.

Direct differentiation at $s=1$ also gives $\partial_s\ceff^F=3/4$ and $\partial_s\ccap^F=-1/2$, yielding the near-transparency expansion in Eq.~\eqref{eq:enhancement}. For the boson, $\partial_s\ceff^B=3/4$ and $\partial_s\ccap^B=1/2$, so both coefficients decrease with reflection, in contrast to the fermion.

The opposite behavior of the bosonic capacity has a simple origin in the single-mode weights. A bosonic mode of entanglement energy $2\omega$ contributes $\omega^2/\sinh^2\omega$ to $\ccap$, which decreases monotonically from its maximal value $1$ at $\omega=0$, whereas a fermionic mode contributes $\omega^2/\cosh^2\omega$, which vanishes at $\omega=0$ and peaks at $\omega\simeq1.2$. Since $\omega_{B,F}\geq x$ for $s<1$, reflection raises all entanglement energies; for bosons this can only lower the variance, while for fermions it moves the low-lying modes toward the peak and can raise it. The fermionic capacity attains its maximum $\ccap/c \simeq 1.09754$ at $s \simeq 0.8029$.

The Dirac fermion requires no separate calculation. It contains two Majorana fermions, $\psi=(\chi_1+i\chi_2)/\sqrt2$, and the energy-independent scattering matrix reads
\begin{equation}
\begin{pmatrix}\psi_L\\ \bar\psi_R\end{pmatrix}=\begin{pmatrix}i\sin\chi & \cos\chi\\ \cos\chi & i\sin\chi\end{pmatrix}\begin{pmatrix}\bar\psi_L\\ \psi_R\end{pmatrix},\qquad s=\cos\chi,
\label{sm:eq:diracS}
\end{equation}
After a constant chiral rephasing, this matrix is real and acts
identically on the two Majorana fields~\cite{Meineri2020}. The reduced density matrix is then a tensor product of the two Majorana ones, so entropy and capacity are additive; equivalently, the single-particle spectrum is $2\omega_F$ with every level doubly degenerate, and Eq.~\eqref{sm:eq:realreplica} holds with $\pi^{-2}\to2\pi^{-2}$. Hence
\begin{equation}
\ceff^{\rm Dirac}(s)=2\,\ceff^{\rm Maj}(s),\qquad \ccap^{\rm Dirac}(s)=2\,\ccap^{\rm Maj}(s),
\label{sm:eq:diracdoubling}
\end{equation}
so the normalized coefficients $\ceff/c$, $\ccap/c$ and $c_{LR}/c=s^2$ are the same functions of $s$ for the two theories. The bound for the Dirac interface therefore follows from the Majorana proof, and the fermionic curve in Fig.~\ref{fig:bound} applies to both interfaces.

\subsection{Comparison and exact sum rules}
\label{sm:sec:sumrules}

Although the fermionic and bosonic families are different interfaces, their coefficients are related by remarkably simple identities. Peschel and Eisler observed that the second derivatives of the two entropy integrals are equal and opposite~\cite{Peschel2012}, so that their sum is linear in $s$. We find that the same structure holds one order higher for the capacity, with a constant rather than vanishing second derivative:
\begin{equation}
\ceff^{B}(s)+\ceff^{\rm Maj}(s)=\tfrac32\,s,\qquad
\ccap^{B}(s)+\ccap^{\rm Maj}(s)=\tfrac32\,s\,(2-s)=\tfrac32\bigl[1-(1-s)^2\bigr].
\label{sm:eq:sumrules}
\end{equation}
Both hold for all $0\leq s\leq1$, with $\tfrac32=c_B+c_{\rm Maj}$; we have verified them numerically from Eq.~\eqref{sm:eq:realreplica}. At $s=1$ both reduce to $c_B+c_{\rm Maj}$, as they must. 

The comparison also yields a free-field analogue of Table~\ref{tab:independence}. Both families have $c_{LR}/c=s^2$, so at equal $s$ they transmit energy identically while differing in $\ceff$ and $\ccap$. Conversely one may tune the two transmissions so that the \emph{entropy} coefficients agree; the capacities then still differ, as shown in Table~\ref{sm:tab:samebutdifferent}. Unlike the holographic pair of Table~\ref{tab:independence}, here $c_{LR}$ differs as well, so the free-field example demonstrates the independence of $\ccap$ from $\ceff$.

\begin{table}[htbp]
\centering
\caption{Free-field interfaces with identical entropy coefficients and different capacities. Each row pair has the same $\ceff/c$ but different $c_{LR}/c$ and $\ccap/c$. The fermionic values apply to both the Majorana and the Dirac interface.}
\label{sm:tab:samebutdifferent}
\begin{tabular}{llccc}
\toprule
Interface & $s$ & $\ceff/c$ & $c_{LR}/c$ & $\ccap/c$\\
\midrule
fermion & 0.6000 & 0.45647 & 0.36000 & 0.99977\\
boson   & 0.3766 & 0.45647 & 0.14183 & 0.57255\\
\midrule
fermion & 0.7000 & 0.58049 & 0.49000 & 1.07194\\
boson   & 0.5017 & 0.58049 & 0.25165 & 0.68423\\
\midrule
fermion & 0.8500 & 0.78220 & 0.72250 & 1.09207\\
boson   & 0.7263 & 0.78220 & 0.52758 & 0.84602\\
\bottomrule
\end{tabular}
\end{table}

\section{Defect perturbation theory}
\label{sm:sec:perturbation}

\subsection{Replica partition function}

We perturb the identity interface of a CFT with central charge $c$ by a normalized spinless primary $\phi$ with $h=\bar h=(1-\varepsilon)/2$, and let $g$ be its dimensionless renormalized coupling at a nearby conformal fixed point. We retain the $O(g^2)$ logarithmic contribution at leading order in $\varepsilon$; $\varepsilon=0$ includes exactly marginal defect deformations.

The vanishing one-point function eliminates the $O(g)$ term. At $O(g^2)$, uniformization maps the interface to $2n$ lines. Integrating the two-point function and subtracting the one-copy contribution gives the change $\delta F(n)$ relative to the identity interface~\cite{Brehm2021}
\begin{equation}
\delta F(n)=2\pi g^2\log\frac{l}{\epsilon}\sum_{j=1}^{n-1}\frac{1-j/n}{\sin(\pi j/n)}=\pi g^2\log\frac{l}{\epsilon}\sum_{j=1}^{n-1}\csc\frac{\pi j}{n},
\label{sm:eq:integersum}
\end{equation}
where the second expression pairs $j$ with $n-j$. Local counterterms for coincident insertions cancel in $F(n)=\log Z_n-n\log Z_1$.

To continue the replica sum, use
\begin{equation}
\csc\frac{\pi j}{n}=\frac2\pi\int_0^\infty\dd u\,\frac{\cosh[(1-2j/n)u]}{\cosh u},\qquad \sum_{j=1}^{n-1}\cosh[(1-2j/n)u]=\frac{\sinh[(n-1)u/n]}{\sinh(u/n)}.
\end{equation}
Setting $u=nx$ gives a convergent integral for real $n>0$,
\begin{equation}
\delta F(n)=2ng^2\log\frac{l}{\epsilon}\int_0^\infty\dd x\,\frac{\sinh[(n-1)x]}{\sinh x\cosh(nx)}=2ng^2\log\frac{l}{\epsilon}\int_0^\infty\dd x\,[\coth x\tanh(nx)-1].
\label{sm:eq:realnperturbation}
\end{equation}
For free fermions, this agrees with the expansion $-2\pi^2g^2\partial_s F_F(n;s)|_{s=1}$ of the exact result in Eq.~\eqref{sm:eq:realreplica}.

\subsection{Entropy and capacity}

Differentiation at $n=1$ yields
\begin{align}
\delta F'(1)&=2g^2\log\frac{l}{\epsilon}\int_0^\infty\dd x\,\frac{x}{\sinh x\cosh x}=\frac{\pi^2}{4}g^2\log\frac{l}{\epsilon},\label{sm:eq:bprime}\\
\delta F''(1)&=4g^2\log\frac{l}{\epsilon}\int_0^\infty\dd x\left[\frac{x}{\sinh x\cosh x}-x^2\sech^2x\right]=\frac{\pi^2}{6}g^2\log\frac{l}{\epsilon}.
\label{sm:eq:bsecond}
\end{align}
The first derivative analytically reproduces the entropy correction inferred numerically in Ref.~\cite{Brehm2021}, while the second establishes an increase in capacity:
\begin{align}
\ceff&=c-\frac{3\pi^2}{2}g^2+O(g^3,\varepsilon g^2),&\ccap&=c+\pi^2g^2+O(g^3,\varepsilon g^2),\label{sm:eq:perturbativecoefficients}\\
\ccap-\ceff&=\frac{5\pi^2}{2}g^2+O(g^3,\varepsilon g^2).
\label{sm:eq:perturbativegap}
\end{align}
Using $\mathcal R=2\pi^2g^2/c+O(g^3,\varepsilon g^2)$~\cite{Brehm2021}, we obtain
\begin{equation}
\frac{\ceff}{c}=1-\frac34\mathcal R+\cdots,\qquad \frac{\ccap}{c}=1+\frac12\mathcal R+\cdots,\qquad \frac{\ccap-\ceff}{c}=\frac54\mathcal R+\cdots.
\label{sm:eq:perturbativeR}
\end{equation}
These slopes agree with the free-fermion result, since $\mathcal R=1-s^2=2(1-s)+\cdots$. For several nearly marginal spinless primaries, replace $g^2$ by $g^aG_{ab}g^b$, where $G_{ab}$ is their two-point matrix. Positivity of $G_{ab}$ in a unitary CFT implies $\ceff<\ccap$ for a nonzero perturbation at this order.

\section{Double-brane holographic interfaces}
\label{sm:sec:doublebrane}

We compute entropy in the $n=1$ geometry and capacity by differentiating the cosmic-brane length with respect to $n$ at $n=1$, allowing both the bulk geometry and the interface branes to vary with $n$.

\subsection{Geometry and replica conditions}

At $n=1$, two pure-tension branes join three locally AdS$_3$ regions with radii $L_L,L_C,L_R$ and a common Newton constant $G_N$. Both branes end on the same boundary interface, which joins two CFTs at $x=0$. We take $A$ to be the right half-line and work at classical order. We choose the central AdS radius to be smaller than both exterior radii. The tension bounds then follow from the Israel junction conditions and the requirement that each brane have an AdS$_2$ induced metric~\cite{Baig2022}:
\begin{equation}
0<L_C<\min(L_L,L_R),\qquad \frac1{L_C}-\frac1{L_i}<8\pi G_N\sigma_i<\frac1{L_C}+\frac1{L_i},\qquad i=L,R.
\label{sm:eq:doubleallowed}
\end{equation}

The boundary map $u+i\tau=\log\zeta$, with $\zeta=x+it$ the Euclidean boundary coordinate, gives two angular strips of width $\pi$, separated at $\tau=\pm\pi/2$. These two lines are the images of the positive- and negative-time halves of the original interface. The central bulk region has zero boundary angular width and introduces no additional boundary CFT. The cutoffs $\epsilon\leq|\zeta|\leq l$ give $\log\epsilon\leq u\leq\log l$, an interval of length $\log\frac{l}{\epsilon}$.

\begin{figure}[t]
\centering
\includegraphics[width=0.95\textwidth]{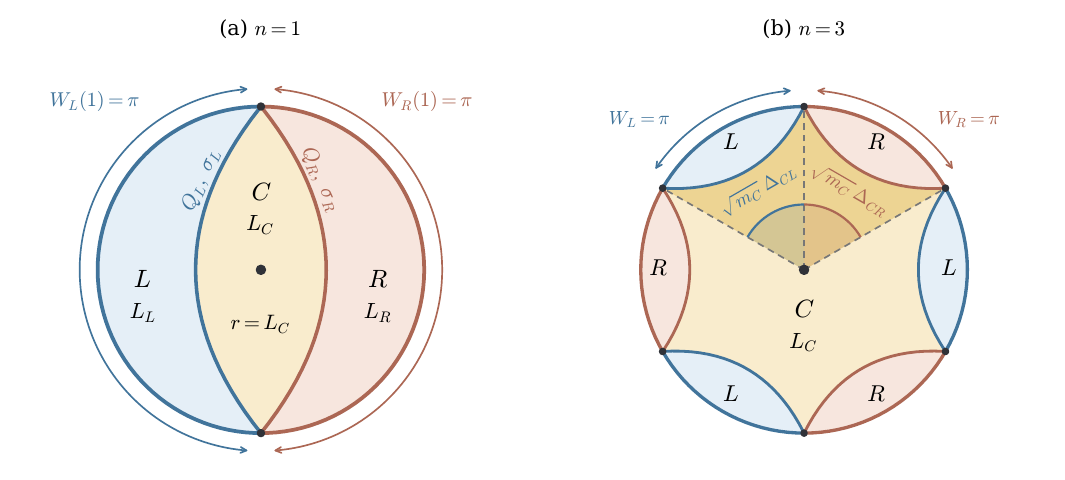}
\caption{Constant-$u$ slices for (a) $n=1$ and (b) $n=3$ (schematic). Each region $j=L,C,R$ has AdS radius $L_j$ and mass parameter $m_j$; the brane $Q_i$ has tension $\sigma_i$. The highlighted sector in (b) represents the replica quotient with opening angle $2\pi/n=2\pi/3$. The exterior widths $W_L=W_R=\pi$ are measured in $\tau_L$ and $\tau_R$, respectively, while the inner arcs show $\sqrt{m_C}\Delta_{CL}$ and $\sqrt{m_C}\Delta_{CR}$, with $\Delta_{Ci}$ measured in $\tau_C$.}
\label{sm:fig:replicaslice}
\end{figure}

For integer $n$, the smooth replica geometry contains $n$ copies of each exterior region and of each interface brane, joined to one central region (Fig.~\ref{sm:fig:replicaslice}). We work in the quotient by replica symmetry, whose boundary retains the two strips of width $\pi$. Near $n=1$, the bulk geometry and brane embeddings vary with $n$ at fixed radii and tensions. We parametrize each region by a dimensionless mass parameter $m_j(n)>0$, $j=L,C,R$, with $m_j(1)=1$ in the original vacuum geometry. In cylindrical coordinates, the metric is~\cite{Bachas2021}
\begin{equation}
\dd s_j^2=\frac{L_j^2\dd r^2}{r^2-L_j^2m_j(n)}+\bigl(r^2-L_j^2m_j(n)\bigr)\dd\tau_j^2+r^2\dd u^2,\qquad r\geq L_j\sqrt{m_j(n)}.
\label{sm:eq:doublemetric}
\end{equation}
Across each brane, we identify points with the same $u$ and $r=\sqrt{g_{uu}}$, while the angular coordinate $\tau_j$ is defined separately in each AdS region. The central region retains the side of each brane containing its axis; each exterior retains the side attached to its boundary strip.

For brane $i=L,R$, let $W_i(n)$ be the exterior boundary strip width measured in $\tau_i$, and $\Delta_{Ci}(n)$ the separation of the brane endpoints measured in $\tau_C$. These quantities refer to the same endpoints but use different angular coordinates. To convert them to polar angles, write $r=L_j\sqrt{m_j(n)}\cosh(\rho/L_j)$, where $\rho$ is proper radial distance from the axis. At fixed $u$, Eq.~\eqref{sm:eq:doublemetric} reduces near $\rho=0$ to
\begin{equation}
\dd s_j^2\simeq\dd\rho^2+m_j(n)\rho^2\dd\tau_j^2.
\end{equation}
The local polar angle is therefore $\sqrt{m_j(n)}\tau_j$, so the central opening angle associated with brane $i$ is $\sqrt{m_C(n)}\Delta_{Ci}(n)$.

In the quotient, a retained exterior axis is smooth, with $\tau_i$ period $2\pi/\sqrt{m_i(n)}$, while the central axis has opening angle $2\pi/n$~\cite{Dong2016}. The exterior boundary widths remain $\pi$, and the two central opening angles must sum to $2\pi/n$. These requirements give
\begin{equation}
W_L(n)=\pi,\qquad W_R(n)=\pi,\qquad \sqrt{m_C(n)}\bigl[\Delta_{CL}(n)+\Delta_{CR}(n)\bigr]=\frac{2\pi}{n}.
\label{sm:eq:doubleclosure}
\end{equation}
The Israel junction conditions determine the brane profiles and hence express $W_i(n)$ and $\Delta_{Ci}(n)$ in terms of the masses. The three conditions in Eq.~\eqref{sm:eq:doubleclosure} therefore determine $m_L(n),m_C(n),m_R(n)$.

At $n=1$, Eq.~\eqref{sm:eq:doublemetric} gives $\dd s\geq r|\dd u|\geq L_C|\dd u|$ throughout the geometry, since $m_j(1)=1$ and $L_C<L_L,L_R$. The central axis at $r=L_C$ saturates this bound and therefore gives the Ryu--Takayanagi surface for the half-line~\cite{Ryu2006}. Its length is $L_C\log\frac{l}{\epsilon}$ to leading logarithmic order. Dividing by $4G_N$ and comparing with the normalization in Eq.~\eqref{sm:eq:coefficientnormalization} gives
\begin{equation}
\ceff=\frac{3L_C}{2G_N}.
\label{sm:eq:doubleentropy}
\end{equation}
In the replica quotient at index $n$, the cosmic brane lies on the central axis at $r=L_C\sqrt{m_C(n)}$ and has length $L_C\sqrt{m_C(n)}\,\log\frac{l}{\epsilon}$. The cosmic-brane formula~\cite{Dong2016} and its derivative at $n=1$ give
\begin{equation}
F(n)-nF'(n)=\frac{L_C\sqrt{m_C(n)}}{4G_N}\log\frac{l}{\epsilon}+o\!\left(\log\frac{l}{\epsilon}\right),\qquad
\ccap=-\frac{3L_C}{4G_N}m_C'(1).
\label{sm:eq:doublecosmic}
\end{equation}
Primes denote derivatives with respect to $n$. Thus capacity requires only $m_C'(1)$.

\subsection{Calculation of capacity}

To determine $m_C'(1)$, we first vary $m_L,m_C,m_R$ independently about $m_j=1$, keeping the AdS radii and brane tensions fixed. To first order, the widths and opening angles change as
\begin{equation}
\begin{aligned}
\delta W_i&=-\frac\pi2\delta m_i+\frac{\delta m_C-\delta m_i}{16G_N\sigma_iL_i},\\
\delta\!\left(\sqrt{m_C}\Delta_{Ci}\right)&=\frac{\delta m_C-\delta m_i}{16G_N\sigma_iL_C}.
\end{aligned}
\label{sm:eq:doublewidthresponse}
\end{equation}
These variations are derived in Sec.~\ref{sm:sec:doublewidths}. When $n$ changes by $\delta n$, we have $\delta m_j=m_j'(1)\delta n$. Differentiating $W_L(n)=W_R(n)=\pi$ at $n=1$ gives the first two equations below. Differentiating the last condition in Eq.~\eqref{sm:eq:doubleclosure} and using these two equations gives the third:
\begin{equation}
\begin{aligned}
\bigl(1+8\pi G_NL_L\sigma_L\bigr)m_L'(1)&=m_C'(1),\\
\bigl(1+8\pi G_NL_R\sigma_R\bigr)m_R'(1)&=m_C'(1),\\
L_Lm_L'(1)+L_Rm_R'(1)&=-4L_C.
\end{aligned}
\label{sm:eq:doublelinearsystem}
\end{equation}
Solving these equations for $m_C'(1)$ gives
\begin{equation}
m_C'(1)=-\frac{4L_C}{\displaystyle\frac{L_L}{1+8\pi G_NL_L\sigma_L}+\displaystyle\frac{L_R}{1+8\pi G_NL_R\sigma_R}},
\label{sm:eq:doublemassresponse}
\end{equation}
and substituting this result into Eq.~\eqref{sm:eq:doublecosmic} gives
\begin{equation}
\ccap=\ceff\,\frac{2L_C}{\displaystyle\frac{L_L}{1+8\pi G_NL_L\sigma_L}+\displaystyle\frac{L_R}{1+8\pi G_NL_R\sigma_R}}.
\label{sm:eq:doublegeneral}
\end{equation}
The lower tension bounds in Eq.~\eqref{sm:eq:doubleallowed} give $L_i/(1+8\pi G_NL_i\sigma_i)<L_C$, proving $\ceff<\ccap$. Equality is approached only when both tensions approach their lower endpoints.

For $L_L=L_R=L$ and $\sigma_L=\sigma_R=\sigma$,
\begin{equation}
\ccap=\ceff\,L_C\left(\frac1L+8\pi G_N\sigma\right),\qquad \ccap c_{LR}=\ceff^2,
\label{sm:eq:doublesymmetric}
\end{equation}
where $c_{LR}$ is given in Eq.~\eqref{eq:doubletransmission}. With unequal tensions, $c_{LR}$ depends only on their sum while capacity also depends on their distribution, as illustrated in Table~\ref{tab:independence}.

\subsection{Variations of the widths and opening angles}
\label{sm:sec:doublewidths}

For the brane between $C$ and $i$, let $k_i$ and $k_{Ci}$ be the $u$-direction eigenvalues of the extrinsic curvature with outward normals. Before expanding about $n=1$, the Israel junction conditions give~\cite{Israel1966,Bachas2021}
\begin{equation}
\begin{gathered}
k_i(r)=k_i^{(0)}+\frac{q_i}{r^2},\qquad
k_i^{(0)}=\frac{(8\pi G_N\sigma_i)^2+L_i^{-2}-L_C^{-2}}{16\pi G_N\sigma_i},\qquad
q_i=\frac{m_C-m_i}{16\pi G_N\sigma_i},\\
k_{Ci}(r)=8\pi G_N\sigma_i-k_i(r).
\end{gathered}
\label{sm:eq:doubleki}
\end{equation}
The induced metric is
\begin{equation}
\dd s_{Q_i}^2=\frac{\dd r^2}{U_i(r)}+r^2\dd u^2,\qquad U_i(r)=\frac{r^2-L_j^2m_j}{L_j^2}-r^2 k_{ji}(r)^2,\qquad j=C,i,
\label{sm:eq:doubleU}
\end{equation}
where $k_{ii}=k_i$. Integrating the embedding from its turning point $r_{0i}$, the largest zero of $U_i$, gives the signed separation of its endpoints:
\begin{equation}
\Delta_{ji}=2\int_{r_{0i}}^\infty\frac{L_jr k_{ji}(r)\,\dd r}{(r^2-L_j^2m_j)\sqrt{U_i(r)}},\qquad j=C,i,
\label{sm:eq:doubleangles}
\end{equation}
with $\Delta_{ii}\equiv\Delta_i$. The retained exterior strip has width~\cite{Bachas2021}
\begin{equation}
W_i=\begin{cases}
2\pi/\sqrt{m_i}-\Delta_i,&\text{if the exterior axis is retained},\\
-\Delta_i,&\text{otherwise}.
\end{cases}
\label{sm:eq:doubleexterior}
\end{equation}

We now evaluate the derivatives at $n=1$, where $m_j=1$ and $q_i=0$. Setting $t=\sqrt{U_i(r)}$ fixes the integration interval to $[0,\infty)$, accounting for the motion of the turning point when differentiating. This gives
\begin{equation}
\begin{aligned}
\left.\frac{\partial\Delta_i}{\partial m_i}\right|_{m_i=1,q_i=0}&=-\frac\pi2\operatorname{sgn}\!\left(k_i^{(0)}\right),\\
\left.\frac{\partial\Delta_i}{\partial q_i}\right|_{m_i=1,q_i=0}&=-\frac{2}{L_i}\int_0^\infty\left[\frac{1}{1+t^2}+\frac{\dd}{\dd t}\frac{t}{t^2+L_i^2(k_i^{(0)})^2}\right]\dd t=-\frac\pi{L_i}.
\end{aligned}
\label{sm:eq:doublelinearangles}
\end{equation}
Differentiating either expression for $W_i$ in Eq.~\eqref{sm:eq:doubleexterior} gives the same $\delta W_i$, whether or not the exterior region contains its axis. On the central side, $k_{Ci}>0$ and $\Delta_{Ci}=\pi$ at $n=1$, and the $1/r^2$ coefficient in $k_{Ci}(r)$ is $-q_i$. Applying Eq.~\eqref{sm:eq:doublelinearangles} on each side therefore gives
\begin{equation}
\delta W_i=-\frac\pi2\delta m_i+\frac\pi{L_i}\delta q_i,\qquad
\delta\!\left(\sqrt{m_C}\Delta_{Ci}\right)=\frac\pi{L_C}\delta q_i.
\label{sm:eq:doubleclosurecheck}
\end{equation}
Substituting $q_i$ from Eq.~\eqref{sm:eq:doubleki} proves Eq.~\eqref{sm:eq:doublewidthresponse}. The case $k_i^{(0)}=0$ follows by continuity of $W_i$.

\section{Lattice models and fitting procedure}
\label{sm:sec:lattice}

\subsection{Free models}

The first two lattice models of the main text are quadratic in fermions, so the reduced density matrix of a block $A$ is Gaussian and is determined by the correlation matrix $\mathcal C_{ij}=\langle c^\dagger_ic_j\rangle$ restricted to $i,j\in A$~\cite{Peschel2003}. Its eigenvalues $\nu_k\in[0,1]$ are the occupations of the single-particle modes of $K_A$, which is itself quadratic, $K_A=\sum_k\varepsilon_kf^\dagger_kf_k+\mathrm{const}$ with $\varepsilon_k=\log[(1-\nu_k)/\nu_k]$. Since the modes are independent, means and variances add, which gives Eq.~\eqref{eq:gaussian}. The $\varepsilon_k$ equal $2\omega_F$ in the notation of Eq.~\eqref{sm:eq:realreplica}, and the density of modes is $\pi^{-2}\log L$ per unit $x$ for a Majorana chain and twice that for a Dirac chain.

\emph{Ising chain with a defect bond.} We take $H=-\sum_{j=1}^{L-1}J_j\sigma^z_j\sigma^z_{j+1}-\sum_{j=1}^L\sigma^x_j$ with $J_j=1$ except $J_{L/2}=\lambda$, and open boundaries. After a Jordan--Wigner transformation this is a chain of $2L$ Majorana modes with uniform nearest-neighbor coupling except for the single bond $(L,L+1)$, whose coupling is $\lambda$. Its transmission amplitude at zero energy is $s=2\lambda/(1+\lambda^2)$~\cite{Eisler2010}. The block $A$ consists of the spins $L/2+1,\ldots,L$, i.e., the Majorana modes $L+1,\ldots,2L$. Since each fermionic mode of $K_A$ is built from two Majorana modes, the sums in Eq.~\eqref{eq:gaussian} carry a factor $1/2$ when written over all eigenvalues of the Majorana covariance matrix, which come in pairs $\nu,1-\nu$.

\emph{Dirac fermion chain with a conformal interface.} We take $H=-\sum_{j=1}^{L-1}t_j(c^\dagger_jc_{j+1}+\mathrm{h.c.})+\sum_j\mu_jc^\dagger_jc_j$ at half filling, with $t_j=1$ except $t_{L/2}=s$, and $\mu_{L/2}=+\sqrt{1-s^2}$, $\mu_{L/2+1}=-\sqrt{1-s^2}$, all other $\mu_j=0$. A bond defect alone is scale-free only at the Fermi point; the two opposite on-site potentials make the single-particle scattering matrix independent of energy over the whole band, so the interface is conformal on the lattice and its transmission amplitude is $s$~\cite{Peschel2012}. The block $A$ is the right half of the chain.

We have checked that, after division by $c$, the two lattice constructions agree with each other at equal $s$, as required by Eq.~\eqref{sm:eq:diracdoubling}, and that the Ising chain reproduces the clean values $\ceff=\ccap=1/2$ at $\lambda=1$.

\subsection{Fitting procedure}

We use $L\in\{64,\ldots,2048\}$ (Ising) and $L\in\{128,\ldots,4096\}$ (Dirac), with $L$ a multiple of four so that the parity of the block length is fixed and the oscillating factor of the lattice corrections has a constant sign~\cite{Cardy2010}.

\emph{Entropy.} The leading lattice correction to $S_A$ is a power of $L$. Extracting its exponent from the local logarithmic slopes of the exact clean-chain data gives $1.00\pm0.01$: the ratio of successive slope errors under $L\to\sqrt2L$ is $0.708(3)$, to be compared with $1/\sqrt2=0.70711$ for an $L^{-1}$ correction. We therefore fit
\begin{equation}
S_A(L)=\frac{\ceff}{6}\log L+a+\frac{b}{L}.
\label{sm:eq:fitS}
\end{equation}

\begin{figure}[t]
\centering
\includegraphics[width=0.92\textwidth]{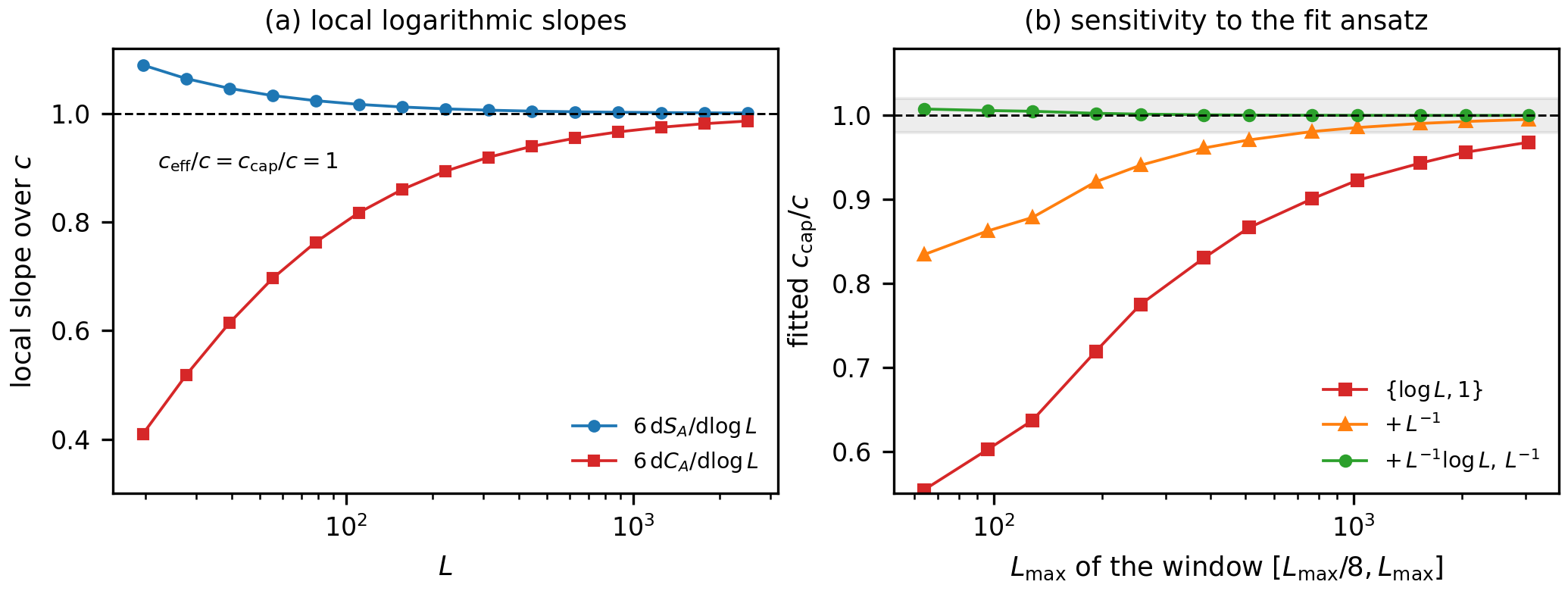}
\caption{Convergence of the lattice coefficients for the clean critical Ising chain, where $\ceff/c=\ccap/c=1$ exactly. (a) Local logarithmic slopes. The entropy converges rapidly; the capacity approaches its asymptote from below and is still $2\%$ low at $L\simeq2000$, since its leading correction is $L^{-1}\log L$ rather than $L^{-1}$. (b) Capacity coefficient $\ccap/c$ extracted from a sliding window $[L_{\max}/8,L_{\max}]$ with three ansatzes: $\{\log L,1\}$ is the bare fit $C_A=\frac{\ccap}{6}\log L+a'$; $+L^{-1}$ adds the correction $b''/L$; and $+L^{-1}\log L,\,L^{-1}$ adds both $b''/L$ and $b'L^{-1}\log L$, i.e., the full form of Eq.~\eqref{sm:eq:fitC}. Only the last is both accurate and stable; the shaded band marks $\pm2\%$.}
\label{fig:convergence}
\end{figure}

\emph{Capacity.} The same correction is not sufficient, for a structural reason. The R\'enyi entropies carry corrections whose exponent depends on the replica index~\cite{Cardy2010},
\begin{equation}
S^{(n)}_A(L)=\frac{c^{(n)}_{\rm eff}}{12}\Bigl(1+\frac1n\Bigr)\log L+a_n+b_n\,L^{-\kappa/n}+\cdots,
\label{sm:eq:renyicorr}
\end{equation}
with $\kappa=1$ for the models at hand. Because $C_A=-2\,\partial_nS^{(n)}_A|_{n=1}$, the derivative acts on the exponent as well as on the amplitude,
\begin{equation}
\partial_n\bigl[b_nL^{-\kappa/n}\bigr]_{n=1}=b_1\kappa\,\frac{\log L}{L^{\kappa}}+b_1'\,\frac{1}{L^{\kappa}},
\label{sm:eq:whylogL}
\end{equation}
so the capacity inherits \emph{both} an $L^{-1}\log L$ and an $L^{-1}$ term. This leads to the fit form:
\begin{equation}
C_A(L)=\frac{\ccap}{6}\log L+a'+\frac{b'\log L+b''}{L}.
\label{sm:eq:fitC}
\end{equation}
In the following fitting process, we treat $\ccap,a',b'$ and $b''$ as some free fitting parameters. The presence of the $L^{-1}\log L$ term is confirmed directly. For such a correction the ratio of successive local-slope errors under $L\to\sqrt2L$ is $\frac{1}{\sqrt2}(\log L-1+\log\sqrt2)/(\log L-1)$, which the exact data follow to within one percent over $L=50$--$2000$, whereas a pure $L^{-1}$ correction would give the constant $0.7071$. The extra factor of $\log L$ is also what makes the capacity converge much more slowly than the entropy: Fig.~\ref{fig:convergence}(a) shows that the local slope of $S_A$ has settled by $L\simeq200$, while that of $C_A$ is still $2\%$ low at $L=2000$.

Table~\ref{sm:tab:fitrobust} compares four ansatzes on exactly solvable data, for the clean chain and for a defect at $s=0.8$ (here $s$ refers to the transmission amplitude for the conformal interface), over several fitting windows. The bare two-parameter fit is strongly biased and its bias depends on the window, so it also distorts the shape of $\ccap(s)$: at $L\leq256$ it underestimates $\ccap$ by $17\%$ at $s=1$ but only by $4\%$ at $s=0.8$. Adding $L^{-1}$ alone removes most but not all of the bias. For the larger fitting windows, including both correction terms in Eq.~\eqref{sm:eq:whylogL} yields \(10^{-4}\)-level accuracy and stable estimates: the fitted value changes by less than $10^{-3}$ when the window is moved from $[64,256]$ to $[256,1024]$, whereas the two-parameter value moves by $13\%$. Figure~\ref{fig:convergence}(b) shows the same comparison as a function of the window.

\begin{table}[htbp]
\centering
\caption{Fitted $\ccap/c$ from exactly solvable data for four ansatzes and several windows, with the deviation from the exact value in parentheses. Top: clean chain, $\ccap/c=1$. Bottom: defect bond $\lambda=1/2$, i.e.\ $s=0.8$, $\ccap/c=1.09752$. The last column is Eq.~\eqref{sm:eq:fitC}.}
\label{sm:tab:fitrobust}
\begin{tabular}{lcccc}
\toprule
window & $\{\log L,1\}$ & $+\,L^{-1}$ & $+\,L^{-1}\log L$ & $+\,L^{-1}\log L,\,L^{-1}$\\
\midrule
\multicolumn{5}{c}{clean chain, $s=1$, exact $\ccap/c=1$}\\
$[64,256]$    & $0.8314\ (-16.9\%)$ & $0.9605\ (-3.9\%)$ & $1.0066\ (+0.7\%)$ & $1.0007\ (+0.1\%)$\\
$[128,512]$   & $0.9018\ (-9.8\%)$  & $0.9805\ (-2.0\%)$ & $1.0029\ (+0.3\%)$ & $1.0002\ (+0.0\%)$\\
$[256,1024]$  & $0.9440\ (-5.6\%)$  & $0.9903\ (-1.0\%)$ & $1.0013\ (+0.1\%)$ & $1.0001\ (+0.0\%)$\\
$[128,1536]$  & $0.9317\ (-6.8\%)$  & $0.9877\ (-1.2\%)$ & $1.0017\ (+0.2\%)$ & $1.0001\ (+0.0\%)$\\
\midrule
\multicolumn{5}{c}{defect, $s=0.8$, exact $\ccap/c=1.09752$}\\
$[64,256]$    & $1.0557\ (-3.8\%)$ & $1.0851\ (-1.1\%)$ & $1.0956\ (-0.2\%)$ & $1.0981\ (+0.1\%)$\\
$[128,512]$   & $1.0722\ (-2.3\%)$ & $1.0915\ (-0.6\%)$ & $1.0969\ (-0.1\%)$ & $1.0977\ (+0.0\%)$\\
$[256,1024]$  & $1.0828\ (-1.3\%)$ & $1.0945\ (-0.3\%)$ & $1.0973\ (-0.0\%)$ & $1.0976\ (+0.0\%)$\\
$[128,1536]$  & $1.0797\ (-1.6\%)$ & $1.0937\ (-0.3\%)$ & $1.0972\ (-0.0\%)$ & $1.0976\ (+0.0\%)$\\
\bottomrule
\end{tabular}
\end{table}

The entropy is far less sensitive: over the same windows all four ansatzes give $\ceff$ within $1.5\%$ of the exact value, and Eq.~\eqref{sm:eq:fitS} within $10^{-4}$. Table~\ref{sm:tab:lattice} lists representative lattice coefficients obtained in this way next to the exact values from Eq.~\eqref{sm:eq:realreplica}. Over the full range of $s$, the maximal deviations are $4\times10^{-5}$ for $\ceff/c$ and $2\times10^{-4}$ for $\ccap/c$ in both models.

\begin{table}[htbp]
\centering
\caption{Lattice coefficients divided by the bulk central charge, compared with the exact free-fermion values obtained from Eq.~\eqref{sm:eq:realreplica}. Ising: defect bond $\lambda$, $s=2\lambda/(1+\lambda^2)$, $c=1/2$. Dirac: conformal interface with hopping $s$ and potentials $\pm\sqrt{1-s^2}$, $c=1$.}
\label{sm:tab:lattice}
\begin{tabular}{cc|cc|cc}
\toprule
\multicolumn{6}{c}{Ising chain with a defect bond}\\
$\lambda$ & $s$ & $\ceff/c$ lattice & exact & $\ccap/c$ lattice & exact\\
\midrule
0.16 & 0.3120 & 0.1594 & 0.1594 & 0.5679 & 0.5678\\
0.30 & 0.5505 & 0.3985 & 0.3985 & 0.9478 & 0.9477\\
0.50 & 0.8000 & 0.7130 & 0.7130 & 1.0977 & 1.0975\\
0.85 & 0.9869 & 0.9804 & 0.9805 & 1.0128 & 1.0126\\
\midrule
\multicolumn{6}{c}{Dirac chain with a conformal interface}\\
& $s$ & $\ceff/c$ lattice & exact & $\ccap/c$ lattice & exact\\
\midrule
& 0.30 & 0.1494 & 0.1495 & 0.5441 & 0.5441\\
& 0.50 & 0.3421 & 0.3422 & 0.8843 & 0.8842\\
& 0.70 & 0.5804 & 0.5805 & 1.0721 & 1.0719\\
& 0.90 & 0.8531 & 0.8531 & 1.0743 & 1.0741\\
\bottomrule
\end{tabular}
\end{table}

\subsection{Interacting interfaces}

The same procedure applies to interacting chains once the Schmidt spectrum of a matrix-product ground state replaces the correlation-matrix eigenvalues, with $p_a=\lambda_a^2$ in
\begin{equation}
S_A=-\sum_ap_a\log p_a,\qquad C_A=\sum_ap_a\log^2p_a-S_A^2 .
\label{sm:eq:schmidtSC}
\end{equation}
Because $C_A$ weights every Schmidt value by $\log^2p_a$, it is more sensitive to the bond dimension than $S_A$.

Ground states of open chains are computed by density-matrix renormalization group (DMRG) method~\cite{White1992} using TeNPy~\cite{Hauschild2018}. The maximal bond dimension is increased with $L$, between $300$ and $800$. For interacting chains the exponent of the leading correction is not known a priori, so we use the direct generalization of Eqs.~\eqref{sm:eq:fitS} and \eqref{sm:eq:fitC},
\begin{equation}
\begin{aligned}
S_A(L)&=\frac{\ceff}{6}\log L+a+b\,L^{-\kappa},\\
C_A(L)&=\frac{\ccap}{6}\log L+a'+b'\,L^{-\kappa}\log L+b''L^{-\kappa},
\end{aligned}
\label{sm:eq:fitkappa}
\end{equation}
and test the dependence of the results on $\kappa$, on the fitting window, and on whether $b''$ is included. As a fit-independent check we also examine the local slopes $6\,\dd S_A/\dd\log L$ and $6\,\dd C_A/\dd\log L$. In all free-fermion chains the capacity slope approaches its asymptote from below (see Fig.~\ref{fig:convergence}(a)). Unless stated otherwise, a quoted uncertainty is the largest deviation from a reference fit over the fitting windows and the two forms of the capacity correction ($b''=0$ and $b''\neq0$) at fixed $\kappa$; the dependence on $\kappa$ is reported separately.

\subsubsection*{$W_3$ interface}

\emph{Model.} The spin chain realization of $W_3$ interface~\cite{Antunes2026} reads $H=H_L+H_R+h_b\,Z_{L/2}Z_{L/2+1}$, with
\begin{equation}
\begin{aligned}
H_L&=v\Biggl[-\sum_{j=1}^{L/2-1}Z_jZ_{j+1}-\sum_{j=1}^{L/2}X_j+g_c\sum_{j=1}^{L/2-2}\bigl(X_jZ_{j+1}Z_{j+2}+Z_jZ_{j+1}X_{j+2}\bigr)\Biggr],\\
H_R&=-\sum_{j=L/2+1}^{L-1}Z_jZ_{j+1}-\sum_{j=L/2+1}^{L}X_j,
\end{aligned}
\label{sm:eq:W3}
\end{equation}
so that no term other than the interface bond couples the two halves. The left half is the O'Brien--Fendley chain~\cite{OBrien2018} at its tricritical point $g_c\simeq0.428$. Its velocity is $v_{\rm TCI}\simeq1.627$, obtained from the finite-size spectrum~\cite{Zou2018}, and the factor $v=2/1.627$ matches it to the velocity $2$ of the critical Ising chain on the right. We use $L=200,220,\ldots,500$ and cut at the interface bond in the middle.

After folding, the two CFTs become a single theory with $c=7/10+1/2=6/5$, and the interface becomes one of its conformal boundary states. Antunes and Rong found that its open-string spectrum,
\begin{equation}
Z_{\rm open}=\chi_0+\chi_{1/2}+2\chi_2+\chi_{1/9}+\chi_{7/9}+\chi_{13/9},
\label{sm:eq:W3open}
\end{equation}
cannot be organized into representations of the product of the two Virasoro algebras, but is organized by the $W_3$ algebra of the second $W_3$ minimal model $[su(3)_2\oplus su(3)_1]/su(3)_3$, which also has $c=6/5$~\cite{Antunes2026}. The characters in Eq.~\eqref{sm:eq:W3open} are $W_3$ characters, and the weights $1/9$, $7/9$ and $13/9$ cannot be written as sums of Ising and TCI weights. The interface is therefore distinct from Gaiotto's RG domain wall, whose spectrum is a sum of products of Ising and TCI characters~\cite{Gaiotto2012}. It commutes with the $\mathbb Z_2$ symmetry $\prod_jX_j$ and with the noninvertible Kramers--Wannier duality of the chain. Its spectrum contains a single relevant $\mathbb Z_2$-even interface operator, of dimension $\Delta_+\simeq0.5$, so that one coupling must be tuned; Ref.~\cite{Antunes2026} located $h_b=1.379$ from the crossing of rescaled energy gaps on periodic chains with two interfaces. The universal data relevant here, $\ccap$ and $\ceff$, are not known analytically; the only general constraint is $\ceff\leq\min(c_L,c_R)=1/2$~\cite{Karch2023}.

\begin{figure}[tbp]
\centering
\includegraphics[width=0.95\textwidth]{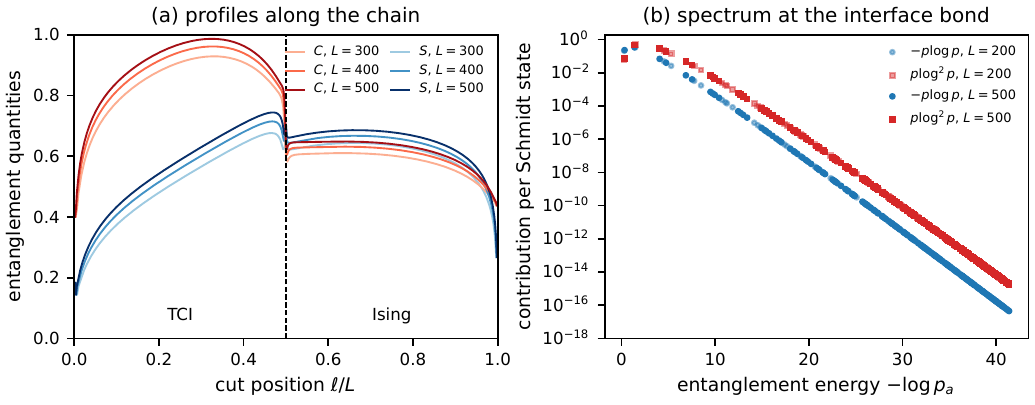}
\caption{$W_3$ interface chain. (a) Capacity (red) and entropy (blue) for every cut position $\ell$, for $L=300$, $400$ and $500$; the dashed line marks the interface bond. The offset between $C$ and $S$ in the TCI half is nonuniversal and also present in the homogeneous TCI chain. (b) Contributions $-p_a\log p_a$ to $S_A$ and $p_a\log^2p_a$ to $\langle K_A^2\rangle$ of the individual Schmidt states at the interface bond, for $L=200$ (light) and $500$ (dark). Both decay exponentially in the entanglement energy $-\log p_a$; the capacity weight decays more slowly but is below $10^{-14}$ at the end of the retained spectrum.}
\label{sm:fig:W3profile}
\end{figure}

The homogeneous O'Brien--Fendley chain at $g_c$ has the same ground state as $H_L$ alone. Fitting its profile to the Calabrese--Cardy form $S(\ell)=\frac{c}{6}\log[\frac{2L}{\pi}\sin\frac{\pi\ell}{L}]+s$ over $10\leq\ell\leq L/4$ gives $c=0.691$ at $L=500$, consistent with $c_{\rm TCI}=7/10$; wider windows give values between $0.66$ and $0.75$. The critical Ising chain is exactly solvable and gives $c=1/2$ (Sec.~\ref{sm:sec:lattice}). The profiles of the $W_3$ chain are shown in Fig.~\ref{sm:fig:W3profile}(a). Both $S$ and $C$ dip at the interface bond. In the TCI half $C$ lies well above $S$; this offset is a nonuniversal constant that is also present in the homogeneous TCI chain, where $S\simeq0.55$ and $C\simeq0.96$ away from the boundaries at $L=500$. Cuts inside either half are not described by the homogeneous formula, because the interface changes the geometry seen by the cut; the bulk theories are therefore checked on homogeneous chains.

The reference fit uses $L\in[200,500]$, $\kappa=1$, the value found for the free-fermion chains, and $b''\neq 0$. We vary the window over $L_{\min}\in\{200,\ldots,280\}$ and $L_{\max}\in\{420,\ldots,500\}$, with at least eight sizes, and include $b''=0$. Over these $50$ fits $\ceff\in[0.4956,0.5011]$ and $\ccap\in[0.5414,0.5461]$, which gives Eq.~\eqref{eq:W3}, and $\ccap-\ceff\in[0.042,0.050]$. Table~\ref{sm:tab:W3} lists representative fitting results.

The correction exponent is not fixed by the data: a free scan of $\kappa$ on $S_A$ does not produce an interior minimum of the residual, indicating slowly varying subleading corrections. Varying $\kappa$ over $[0.8,1.2]$ shifts $\ceff$ by up to $0.007$ and $\ccap$ by up to $0.020$. The capacity is more sensitive because both of its correction terms depend on $\kappa$. Throughout this range $\ccap-\ceff\geq0.035$ and $\ccap>1/2$. The bound $\ceff\leq\min(c_L,c_R)$, which the fitted $\ceff$ exceeds for $\kappa\leq0.9$, disfavors smaller values of $\kappa$. The bare two-parameter fit, which underestimates $\ccap$ strongly for free fermions, gives $\ccap-\ceff=0.007$, illustrating again that the $L^{-\kappa}\log L$ term is essential.

The slopes cross at $L\simeq240$, and at the largest sizes $L\simeq490$ they are $0.481$ for $S_A$ and $0.496$ for $C_A$. The difference between them therefore grows with $L$ and has the sign required by Eq.~\eqref{eq:bound}. The capacity slope has not yet crossed $1/2$, so $\ccap>\min(c_L,c_R)$ relies on the extrapolation. 

\begin{table}[htbp]
\centering
\caption{Coefficients for the $W_3$ interface from Eq.~\eqref{sm:eq:fitkappa}. The reference fit uses $L\in[200,500]$, $\kappa=1$ and $b''\neq0$; each other row changes the indicated choices. The last row is the bare two-parameter fit.}
\label{sm:tab:W3}
\begin{tabular}{lccc}
\toprule
fit & $\ceff$ & $\ccap$ & $\ccap-\ceff$\\
\midrule
reference: $L\in[200,500]$, $\kappa=1$, $b''\neq0$ & $0.4976$ & $0.5459$ & $0.0483$\\
window $[240,500]$ & $0.4995$ & $0.5460$ & $0.0466$\\
window $[280,500]$ & $0.5011$ & $0.5461$ & $0.0451$\\
window $[200,440]$ & $0.4962$ & $0.5458$ & $0.0496$\\
window $[240,460]$ & $0.4985$ & $0.5460$ & $0.0474$\\
$b''=0$ & $0.4976$ & $0.5419$ & $0.0443$\\
\midrule
$\kappa=0.8$ & $0.5045$ & $0.5527$ & $0.0482$\\
$\kappa=0.8$, $b''=0$ & $0.5045$ & $0.5657$ & $0.0612$\\
$\kappa=0.9$ & $0.5007$ & $0.5494$ & $0.0488$\\
$\kappa=1.1$ & $0.4951$ & $0.5424$ & $0.0473$\\
$\kappa=1.2$ & $0.4930$ & $0.5391$ & $0.0461$\\
$\kappa=1.2$, $b''=0$ & $0.4930$ & $0.5284$ & $0.0354$\\
\midrule
two-parameter, $\{\log L,1\}$ & $0.4701$ & $0.4770$ & $0.0070$\\
\bottomrule
\end{tabular}
\end{table}

\subsubsection*{Renormalization-group domain wall}

In this subsection, we take the O'Brien--Fendley chain without velocity matching,
\begin{equation}
H=-\sum_{j=1}^{L-1}Z_jZ_{j+1}-\sum_{j=1}^{L}X_j+\sum_{j=1}^{L-2}g_j\bigl(X_jZ_{j+1}Z_{j+2}+Z_jZ_{j+1}X_{j+2}\bigr),
\label{sm:eq:RGDW}
\end{equation}
with $g_j=g_1$ for $j\leq L/2-1$ and $g_j=g_3=g_c$ for $j\geq L/2$, so that the three-site coupling changes at the center; the cut is between sites $L/2$ and $L/2+1$. For $g_1<g_c$ the left half lies on the critical Ising region. Close to $g_c$ it is the TCI CFT perturbed by $\phi_{1,3}$, of dimension $\Delta=6/5$ and RG eigenvalue $y=2-\Delta=4/5$, with crossover length $\xi\propto(g_c-g_1)^{-5/4}$. At scales beyond $\xi$ the interface then approaches the RG domain wall of Ref.~\cite{Gaiotto2012}, for which $\ceff$ becomes close to $\min(c_L,c_R)=1/2$~\cite{Tang2024}. We use \(\kappa=4/5\) as a phenomenological crossover-motivated ansatz and test the stability of the resulting coefficients, including $b''$, over $200\leq L\leq500$, for different values $0\leq g_1\leq0.40$ and for the uniform chain $g_1=g_c$.

\begin{figure}[tbp]
\centering
\includegraphics[width=0.58\textwidth]{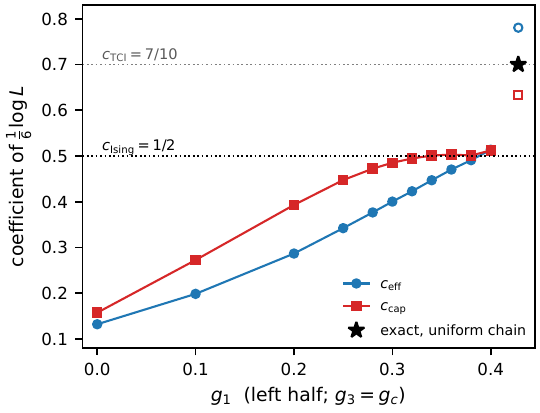}
\caption{Coefficients for the TCI--Ising RG domain wall on the chain of Eq.~\eqref{sm:eq:RGDW}, as functions of the left coupling $g_1$ at fixed $g_3=g_c$. Symbols: Eq.~\eqref{sm:eq:fitkappa} with $\kappa=4/5$, including $b''$, over $200\leq L\leq500$. The chain realizes the RG domain wall for $g_1\lesssim g_c$, where $\ccap\simeq1/2$ and $\ceff$ approaches $1/2$ from below. Open symbols: the uniform chain $g_1=g_c$, with no interface, whose exact value $\ceff=\ccap=7/10$ is marked by the star. Dotted lines mark $c_{\rm Ising}$ and $c_{\rm TCI}$.}
\label{sm:fig:RGDW}
\end{figure}

Figure~\ref{sm:fig:RGDW} shows the result. The two ends of the range of $g_1$ play different roles. For $g_1<g_c$ the left half is on the critical Ising line, but the chain of Eq.~\eqref{sm:eq:RGDW} is not velocity matched: the velocity of the left half varies with $g_1$ and equals that of the TCI half only at $g_1=g_c$. Only for $g_1$ close to $g_c$ is the left half a weak $\phi_{1,3}$ deformation of the TCI chain, with a small velocity mismatch, so that the junction flows to the RG domain wall of Ref.~\cite{Gaiotto2012}. As $g_1$ moves away from $g_c$, both the velocity mismatch and the deviation from the perturbative regime grow, and the lattice junction is no longer guaranteed to realize this interface; the same pair of CFTs admits several conformal interfaces, among them the $W_3$ interface. The coefficients at small $g_1$, where $\ceff$ lies between $0.13$ and $0.40$, far below the value $\min(c_L,c_R)=1/2$, reflect this and are not used to test the bound.

The RG domain wall is therefore realized for $g_1\lesssim g_c$. In this regime, $0.32\leq g_1\leq0.38$, the capacity is $\ccap=0.494$--$0.503$, consistent with $\ccap\simeq1/2$, while the entropy coefficient approaches $1/2$ from below, $\ceff=0.423$--$0.491$; the gap $\ccap-\ceff$ decreases from $0.07$ to $0.01$ as $g_1\to g_c$. The fitted coefficients thus satisfy $\ceff<\ccap$ throughout the regime in which the lattice chain realizes the RG domain wall, supporting Eq.~\eqref{eq:bound}. 

Closer to $g_c$ the crossover length $\xi\propto(g_c-g_1)^{-5/4}$ becomes comparable to the system size, and at fixed $L$ the chain crosses over to the homogeneous TCI chain. At $g_1=0.40$ the fitted $\ceff=0.514$ already exceeds $\min(c_L,c_R)=1/2$, and $\ceff$ and $\ccap$ coincide within $0.002$; at $g_1=g_c$ there is no interface, and the fit approaches the TCI values, giving $\ceff=0.78$ and $\ccap=0.63$ compared with the exact $\ceff=\ccap=7/10$. The deviation at this point indicates the finite-size effects when the TCI region dominates.

\FloatBarrier


\begin{thebibliography}{99}
\bibitem{Oshikawa1996} M. Oshikawa and I. Affleck, Defect lines in the Ising model and boundary states on orbifolds, Phys. Rev. Lett. \textbf{77}, 2604 (1996), \href{https://arxiv.org/abs/hep-th/9606177}{arXiv:hep-th/9606177}.
\bibitem{Oshikawa1997} M. Oshikawa and I. Affleck, Boundary conformal field theory approach to the critical two-dimensional Ising model with a defect line, Nucl. Phys. B \textbf{495}, 533 (1997), \href{https://arxiv.org/abs/cond-mat/9612187}{arXiv:cond-mat/9612187}.
\bibitem{Bachas2002} C. Bachas, J. de Boer, R. Dijkgraaf, and H. Ooguri, Permeable conformal walls and holography, J. High Energy Phys. \textbf{06} (2002) 027, \href{https://arxiv.org/abs/hep-th/0111210}{arXiv:hep-th/0111210}.
\bibitem{Quella2007} T. Quella, I. Runkel, and G. M. T. Watts, Reflection and transmission for conformal defects, J. High Energy Phys. \textbf{04} (2007) 095, \href{https://arxiv.org/abs/hep-th/0611296}{arXiv:hep-th/0611296}.
\bibitem{Sakai2008} K. Sakai and Y. Satoh, Entanglement through conformal interfaces, J. High Energy Phys. \textbf{12} (2008) 001, \href{https://arxiv.org/abs/0809.4548}{arXiv:0809.4548}.
\bibitem{Eisler2010} V. Eisler and I. Peschel, Entanglement in fermionic chains with interface defects, Ann. Phys. (Berlin) \textbf{522}, 679 (2010), \href{https://arxiv.org/abs/1005.2144}{arXiv:1005.2144}.
\bibitem{Peschel2012} I. Peschel and V. Eisler, Exact results for the entanglement across defects in critical chains, J. Phys. A: Math. Theor. \textbf{45}, 155301 (2012), \href{https://arxiv.org/abs/1201.4104}{arXiv:1201.4104}.
\bibitem{Calabrese2012} P. Calabrese, M. Mintchev, and E. Vicari, Entanglement entropy of quantum wire junctions, J. Phys. A: Math. Theor. \textbf{45}, 105206 (2012), \href{https://arxiv.org/abs/1110.5713}{arXiv:1110.5713}.
\bibitem{Brehm2015} E. M. Brehm and I. Brunner, Entanglement entropy through conformal interfaces in the 2D Ising model, J. High Energy Phys. \textbf{09} (2015) 080, \href{https://arxiv.org/abs/1505.02647}{arXiv:1505.02647}.
\bibitem{Gutperle2016Topological} M. Gutperle and J. D. Miller, A note on entanglement entropy for topological interfaces in RCFTs, J. High Energy Phys. \textbf{04} (2016) 176, \href{https://arxiv.org/abs/1512.07241}{arXiv:1512.07241}.
\bibitem{Gutperle2016Holographic} M. Gutperle and J. D. Miller, Entanglement entropy at holographic interfaces, Phys. Rev. D \textbf{93}, 026006 (2016), \href{https://arxiv.org/abs/1511.08955}{arXiv:1511.08955}.
\bibitem{Wen2018} X. Wen, Y. Wang, and S. Ryu, Entanglement evolution across a conformal interface, J. Phys. A: Math. Theor. \textbf{51}, 195004 (2018), \href{https://arxiv.org/abs/1711.02126}{arXiv:1711.02126}.
\bibitem{Rogerson2022} D. Rogerson, F. Pollmann, and A. Roy, Entanglement entropy and negativity in the Ising model with defects, J. High Energy Phys. \textbf{06} (2022) 165, \href{https://arxiv.org/abs/2204.03601}{arXiv:2204.03601}.
\bibitem{Tang2024} Q. Tang, Z. Wei, Y. Tang, X. Wen, and W. Zhu, Universal entanglement signatures of interface conformal field theories, Phys. Rev. B \textbf{109}, L041104 (2024), \href{https://arxiv.org/abs/2308.03646}{arXiv:2308.03646}.
\bibitem{Meineri2020} M. Meineri, J. Penedones, and A. Rousset, Colliders and conformal interfaces, J. High Energy Phys. \textbf{02} (2020) 138, \href{https://arxiv.org/abs/1904.10974}{arXiv:1904.10974}.
\bibitem{Bachas2020} C. Bachas, S. Chapman, D. Ge, and G. Policastro, Energy reflection and transmission at 2D holographic interfaces, Phys. Rev. Lett. \textbf{125}, 231602 (2020), \href{https://arxiv.org/abs/2006.11333}{arXiv:2006.11333}.
\bibitem{Bachas2023Thick} C. Bachas, S. Baiguera, S. Chapman, G. Policastro, and T. Schwartzman, Energy transport for thick holographic branes, Phys. Rev. Lett. \textbf{131}, 021601 (2023), \href{https://arxiv.org/abs/2212.14058}{arXiv:2212.14058}.
\bibitem{Liu2025} Y. Liu and C.-Y. Wang, Energy transport in holographic junctions, J. High Energy Phys. \textbf{10} (2025) 205, \href{https://arxiv.org/abs/2506.19553}{arXiv:2506.19553}.
\bibitem{Barad2025} R. Barad, Q. Tang, and X. Wen, Dissipation meets conformal interface in open quantum systems: How the relaxation rate is suppressed, Phys. Rev. B \textbf{112}, 235143 (2025), \href{https://arxiv.org/abs/2505.04715}{arXiv:2505.04715}.
\bibitem{KarchWang2026} A. Karch and M. Wang, Universality of dissipation across holographic interfaces, \href{https://arxiv.org/abs/2601.16888}{arXiv:2601.16888} (2026).
\bibitem{Barad2026} R. Barad, T. Wu, Q. Tang, and X. Wen, Universal Suppression of Dissipation across Conformal Interface in Open Quantum Critical Systems, \href{https://arxiv.org/abs/2609.19292}{arXiv:2609.19292} (2026).
\bibitem{Yao2010} H. Yao and X.-L. Qi, Entanglement entropy and entanglement spectrum of the Kitaev model, Phys. Rev. Lett. \textbf{105}, 080501 (2010), \href{https://arxiv.org/abs/1001.1165}{arXiv:1001.1165}.
\bibitem{Nakaguchi2016} Y. Nakaguchi and T. Nishioka, A holographic proof of R\'enyi entropic inequalities, J. High Energy Phys. \textbf{12} (2016) 129, \href{https://arxiv.org/abs/1606.08443}{arXiv:1606.08443}.
\bibitem{deBoer2019} J. de Boer, J. J\"arvel\"a, and E. Keski-Vakkuri, Aspects of capacity of entanglement, Phys. Rev. D \textbf{99}, 066012 (2019), \href{https://arxiv.org/abs/1807.07357}{arXiv:1807.07357}.
\bibitem{Kawabata2021} K. Kawabata, T. Nishioka, Y. Okuyama, and K. Watanabe, Replica wormholes and capacity of entanglement, J. High Energy Phys. \textbf{10} (2021) 227, \href{https://arxiv.org/abs/2105.08396}{arXiv:2105.08396}.
\bibitem{Nandy2021} P. Nandy, Capacity of entanglement in local operators, J. High Energy Phys. \textbf{07} (2021) 019, \href{https://arxiv.org/abs/2106.00228}{arXiv:2106.00228}.
\bibitem{Arias2023} R. Arias, G. Di Giulio, E. Keski-Vakkuri, and E. Tonni, Probing RG flows, symmetry resolution and quench dynamics through the capacity of entanglement, J. High Energy Phys. \textbf{03} (2023) 175, \href{https://arxiv.org/abs/2301.02117}{arXiv:2301.02117}.
\bibitem{Datta2015} N. Datta and F. Leditzky, Second-order asymptotics for source coding, dense coding and pure-state entanglement conversions, IEEE Trans. Inf. Theory \textbf{61}, 582 (2015), \href{https://arxiv.org/abs/1403.2543}{arXiv:1403.2543}.
\bibitem{Bao2026} N. Bao and J. March, The capacity of entanglement and holographic entropies at finite resources, \href{https://arxiv.org/abs/2608.26377}{arXiv:2608.26377} (2026).
\bibitem{Calabrese2004} P. Calabrese and J. Cardy, Entanglement entropy and quantum field theory, J. Stat. Mech. (2004) P06002, \href{https://arxiv.org/abs/hep-th/0405152}{arXiv:hep-th/0405152}.
\bibitem{Karch2023} A. Karch, Y. Kusuki, H. Ooguri, H.-Y. Sun, and M. Wang, Universality of effective central charge in interface CFTs, J. High Energy Phys. \textbf{11} (2023) 126, \href{https://arxiv.org/abs/2308.05436}{arXiv:2308.05436}.
\bibitem{Karch2024} A. Karch, Y. Kusuki, H. Ooguri, H.-Y. Sun, and M. Wang, Universal bound on effective central charge and its saturation, Phys. Rev. Lett. \textbf{133}, 091604 (2024), \href{https://arxiv.org/abs/2404.01515}{arXiv:2404.01515}.
\bibitem{SM} See the \hyperref[sec:supplement]{Supplemental Material} included in this manuscript for the free-field R\'enyi formulas and proof, defect perturbation theory, double-brane replica calculations, and for the lattice models, the fitting procedure, and the numerical results for the $W_3$ interface and the RG domain wall.
\bibitem{Brehm2021} E. M. Brehm, Defects and perturbation, J. High Energy Phys. \textbf{04} (2021) 300, \href{https://arxiv.org/abs/2011.10795}{arXiv:2011.10795}.
\bibitem{Ryu2006} S. Ryu and T. Takayanagi, Holographic derivation of entanglement entropy from AdS/CFT, Phys. Rev. Lett. \textbf{96}, 181602 (2006), \href{https://arxiv.org/abs/hep-th/0603001}{arXiv:hep-th/0603001}.
\bibitem{Dong2016} X. Dong, The gravity dual of R\'enyi entropy, Nat. Commun. \textbf{7}, 12472 (2016), \href{https://arxiv.org/abs/1601.06788v3}{arXiv:1601.06788v3}.
\bibitem{Baig2022} S. A. Baig and A. Karch, Double brane holographic model dual to 2d ICFTs, J. High Energy Phys. \textbf{10} (2022) 022, \href{https://arxiv.org/abs/2206.01752v2}{arXiv:2206.01752v2}.
\bibitem{Israel1966} W. Israel, Singular hypersurfaces and thin shells in general relativity, Nuovo Cimento B \textbf{44}, 1 (1966), \href{https://doi.org/10.1007/BF02710419}{doi:10.1007/BF02710419}.
\bibitem{Karch2021} A. Karch, Z.-X. Luo, and H.-Y. Sun, Universal relations for holographic interfaces, J. High Energy Phys. \textbf{09} (2021) 172, \href{https://arxiv.org/abs/2107.02165}{arXiv:2107.02165}.
\bibitem{KarchWang2023} A. Karch and M. Wang, Universal behavior of entanglement entropies in interface CFTs from general holographic spacetimes, J. High Energy Phys. \textbf{06} (2023) 145, \href{https://arxiv.org/abs/2211.09148}{arXiv:2211.09148}.
\bibitem{Eisler2012Evolution} V. Eisler and I. Peschel, On entanglement evolution across defects in critical chains, EPL \textbf{99}, 20001 (2012), \href{https://arxiv.org/abs/1205.4331}{arXiv:1205.4331}.
\bibitem{Capizzi2023} L. Capizzi and V. Eisler, Entanglement evolution after a global quench across a conformal defect, SciPost Phys. \textbf{14}, 070 (2023), \href{https://arxiv.org/abs/2209.03297}{arXiv:2209.03297}.
\bibitem{Peschel2003} I. Peschel, Calculation of reduced density matrices from correlation functions, J. Phys. A: Math. Gen. \textbf{36}, L205 (2003), \href{https://arxiv.org/abs/cond-mat/0212631}{arXiv:cond-mat/0212631}.
\bibitem{Cardy2010} J. Cardy and P. Calabrese, Unusual corrections to scaling in entanglement entropy, J. Stat. Mech. (2010) P04023, \href{https://arxiv.org/abs/1002.4353}{arXiv:1002.4353}.
\bibitem{Cogburn2024} C. V. Cogburn, A. L. Fitzpatrick, and H. Geng, CFT and lattice correlators near an RG domain wall between minimal models, SciPost Phys. Core \textbf{7}, 021 (2024), \href{https://arxiv.org/abs/2308.00737}{arXiv:2308.00737}.
\bibitem{Antunes2026} A. Antunes and J. Rong, A new Ising/tricritical-Ising interface: From $\mathcal W_3$ symmetry to Rydberg atoms, \href{https://arxiv.org/abs/2605.23848}{arXiv:2605.23848} (2026).
\bibitem{Brehm2016} E. M. Brehm, I. Brunner, D. Jaud, and C. Schmidt-Colinet, Entanglement and topological interfaces, Fortschr. Phys. \textbf{64}, 516 (2016), \href{https://arxiv.org/abs/1512.05945}{arXiv:1512.05945}.
\bibitem{Bachas2021} C. Bachas and V. Papadopoulos, Phases of holographic interfaces, J. High Energy Phys. \textbf{04} (2021) 262, \href{https://arxiv.org/abs/2101.12529v3}{arXiv:2101.12529v3}.
\bibitem{OBrien2018} E. O'Brien and P. Fendley, Lattice supersymmetry and order-disorder coexistence in the tricritical Ising model, Phys. Rev. Lett. \textbf{120}, 206403 (2018), \href{https://arxiv.org/abs/1712.06662}{arXiv:1712.06662}.
\bibitem{White1992} S. R. White, Density matrix formulation for quantum renormalization groups, Phys. Rev. Lett. \textbf{69}, 2863 (1992).
\bibitem{Hauschild2018} J. Hauschild and F. Pollmann, Efficient numerical simulations with tensor networks: Tensor Network Python (TeNPy), SciPost Phys. Lect. Notes \textbf{5} (2018), \href{https://arxiv.org/abs/1805.00055}{arXiv:1805.00055}.
\bibitem{Gaiotto2012} D. Gaiotto, Domain walls for two-dimensional renormalization group flows, J. High Energy Phys. \textbf{12} (2012) 103, \href{https://arxiv.org/abs/1201.0767}{arXiv:1201.0767}.
\bibitem{Zou2018} Y. Zou, A. Milsted, and G. Vidal, Conformal data and renormalization group flow in critical quantum spin chains using periodic uniform matrix product states, Phys. Rev. Lett. \textbf{121}, 230402 (2018), \href{https://arxiv.org/abs/1710.05397}{arXiv:1710.05397}.
\bibitem{Furuta2026} Y. Furuta, Y. Kusuki, T. Onagi and Y. Tang, \emph{to appear}.
\bibitem{Kokail2021} C. Kokail, R. van Bijnen, A. Elben, B. Vermersch, and P. Zoller, Entanglement Hamiltonian tomography in quantum simulation, Nat. Phys. \textbf{17}, 936 (2021), \href{https://arxiv.org/abs/2009.09000}{arXiv:2009.09000}.
\bibitem{Joshi2023} M. K. Joshi, C. Kokail, R. van Bijnen, F. Kranzl, T. V. Zache, R. Blatt, C. F. Roos, and P. Zoller, Exploring large-scale entanglement in quantum simulation, Nature \textbf{624}, 539 (2023), \href{https://arxiv.org/abs/2306.00057}{arXiv:2306.00057}.
\end{thebibliography}
\end{document}